%% file: main.tex
\documentclass[USenglish,oneside,twocolumn]{article}

\usepackage[utf8]{inputenc}
\usepackage[big]{dgruyter_NEW}
\usepackage[table]{xcolor}
\usepackage{amsfonts}
\usepackage{amssymb}
\usepackage{amsmath}
\usepackage{subfig}
\usepackage{comment}
\usepackage{svg}
\newif\ifediting
 
\DOI{foobar}

\cclogo{\includegraphics{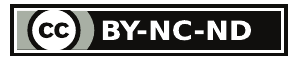}}
 
\def\REMARK#1#2{        
  \ifediting
  \begin{center}
  \noindent\fbox{
  \begin{minipage}[b]{0.45\textwidth}
  \textbf{[#1]: #2}
  \end{minipage}}
  \end{center}
  \fi}

\def\NB#1{\REMARK{Nataliia}{#1}}
\def\IF#1{\REMARK{Imane}{#1}}

\def\SHORTEN{\vspace{-0.5cm}}

\def\domain#1{\texttt{#1}}
\def\etal{et al.}
\hypersetup{draft}
\input{macros}
\begin{document}

   \author*[1]{Imane fouad}

   \author[2]{ Nataliia Bielova}

   \author[3]{Arnaud Legout}

   \author[4]{Natasa Sarafijanovic-Djukic}

\title{\huge Missed by Filter Lists: 
Detecting Unknown Third-Party Trackers with Invisible Pixels}

  \runningtitle{Detecting Unknown Trackers via Invisible Pixels}

\begin{abstract}
{
Web tracking has been extensively studied over the last decade. 
To detect tracking, 
previous studies and user tools rely on  
filter lists. 
However, 
it has been shown that filter lists miss trackers.
In this paper, we propose an alternative method 
to detect trackers inspired by analyzing behavior of invisible pixels. 
By crawling 84,658 webpages from 8,744 domains, we detect that third-party invisible pixels are widely 
deployed: they are present on more than 94.51\% of domains and constitute 35.66\% of all third-party images. 
We 
propose a fine-grained behavioral classification of tracking  based on the analysis of 
invisible pixels. We use this classification to detect new categories of tracking and uncover 
new collaborations between domains on the full dataset of 
$4,216,454$
third-party requests.    
We demonstrate that 
two popular methods to detect tracking, based on EasyList\&EasyPrivacy 
and on Disconnect lists 
respectively miss 25.22\% and 30.34\%  of the trackers that we detect.
Moreover, we find that 
if we combine 
all three lists, $379,245$  requests 
originated from 8,744 domains still 
 track users on 68.70\% of websites. 
}

\end{abstract}


  \journalname{Proceedings on Privacy Enhancing Technologies}
\DOI{Editor to enter DOI}
  \startpage{1}
  \received{..}
  \revised{..}
  \accepted{..}

  \journalyear{..}
  \journalvolume{..}
  \journalissue{..}

\maketitle

\input{introduction}

\input{background}
\input{methodology}

\input{results}

\input{comparing}

\input{relatedwork}

\input{consumerlists}

\input{limitation}

\input{conclusion}


\bibliographystyle{plain} 
\bibliography{papers,articles,urls} 

\input{appendix}

\end{document}

%% file: macros.tex
\def\totaldomainssucc{\ensuremath {8,744 }}
\def\totalpagessucc{\ensuremath {84,658}}
\def\totalimages{\ensuremath {2,297,716}}

\def\ourmethod{\ensuremath{\mathsf{BehaviorTrack}}}
\def\id{identifier cookie}

\def\al{et al.}

\def\ELEP{EL\&EP}
\def\DISC{Disconnect}
\def\disc{Disconnect}

%% file: introduction.tex
\section{Introduction}
\label{sec:intro}

The Web has become an essential part of our lives: billions are using Web applications on a daily basis and while doing so, are 
placing \emph{digital traces} on millions of websites. 
Such traces allow advertising companies, as well as data brokers to continuously profit from collecting 
a vast amount of data associated to the users. 
Recent 
works have shown that 
advertising networks and data brokers 
use a wide range of techniques 
to track users across the 
Web 
\cite{Solt-etal-09-Flash,Roes-etal-12-NSDI,Ayen-etal-11-Flash,Olej-etal-14-NDSS,Engl-Nara-16-CCS,
  Ecke-10-PET,Acar-etal-13-CCS,Niki-etal-13-SP,Lern-etal-16-USENIX,Raza-etal-18-NDSS}, 
from standard stateful cookie-based tracking~\cite{Roes-etal-12-NSDI,Engl-etal-15-WWW}, 
to stateless fingerprinting~\cite{Niki-etal-13-SP,Cao-etal-17-NDSS,Engl-Nara-16-CCS,Acar-etal-13-CCS}. 

In the last decade, numerous studies measured prevalence of third-party trackers
on the Web~\cite{Roes-etal-12-NSDI,
Acar-etal-13-CCS,Niki-etal-13-SP,Engl-Nara-16-CCS,Lern-etal-16-USENIX,
Bash-etal-16-USENIX, Bash-Wils-18-PETS,Libe-etal-18-UO,Libe-etal-18-UO-GDPR,
Yu-etal-16-WWW}.
Web Tracking is often considered in the context of targeted behavioral advertising, but it's not limited to ads.  
Third-party tracking has  become deeply integrated into the Web contents that  owners include in their websites.

{\em But what makes a tracker?} How to recognize that a third-party request is performing tracking? 
To detect trackers, 
the research community 
applied a variety 
of methodologies.
The most known Web tracking technique 
is based on \emph{cookies}, 
but only some cookies contain unique identifiers and hence are capable of tracking the users.
Some studies detect trackers 
by analysing 
cookie storage, 
and third-party requests and responses that set or send 
cookies~\cite{Roes-etal-12-NSDI,Lern-etal-16-USENIX},
while other works  measured the mere presence of third-party cookies~\cite{Libe-etal-18-UO,Libe-etal-18-UO-GDPR}. 
To measure \emph{cookie syncing}, 
researchers applied various heuristics to filter cookies with unique identifiers 
\cite{Engl-etal-15-WWW,Acar-etal-14-CCS,Engl-Nara-16-CCS}. However, this approach has never been applied to detect tracking at large scale.
Overall, previous works provide
different methods to 
identify 
third-party requests that are responsible for tracking \cite{Roes-etal-12-NSDI, Yu-etal-16-WWW}.
Detection of identifier cookies and analysing behaviors of third-party domains is a complex 
task. 
Therefore, most of the state-of-the-art works that aim at measuring 
trackers at large scale rely on \emph{filter lists}. 
In particular, EasyList~\cite{easylist} and EasyPrivacy~\cite{easyprivacy} (\ELEP) 
and \DISC~\cite{DisconnectList} lists became the \emph{de facto} approach to detect third-party tracking requests 
in privacy and measurement 
communities
\cite{Engl-Nara-16-CCS,
Bash-etal-16-USENIX,Laui-etal-17-NDSS,Raza-etal-18-NDSS,Ikra-etal-17-PETS,
Engl-etal-18-PETS,Bash-Wils-18-PETS,Bash-etal-18-IMC,Iord-etal-18-IMC}\footnote{We 
summarize the usage of filter lists in security, privacy and web measurement community 
in Table~\ref{tab:ELEPDinliterature} in the Appendix.}.  
 EasyList and EasyPrivacy
  are the most popular publicly maintained blacklist of known advertising and tracking requests, 
used by the popular 
blocking 
extensions AdBlock Plus~\cite{adblockplus}
and uBlockOrigin~\cite{uBlockOrigin}.
Disconnect is another very popular list for detecting domains known for tracking, 
used in Disconnect browser extension~\cite{Disconnect} and in tracking protection of Firefox browser~\cite{firefox}.

Nevertheless, 
filter lists detect only known tracking and ad-related requests. 
 Therefore, 
a tracker can avoid this detection 
by using a different subdomain for tracking,
or  wholly register a new domain if the filter list block the entire domain.
Even though, the second option is quite challenging because in such case, 
all the associated publishers would need to update their pages.
Third parties can also incorporate tracking behavior into functional website content, 
which is never blocked by filter lists because blocking functional content would harm user experience.
Therefore, it is interesting to evaluate how effective are filter lists at detecting trackers, how 
many trackers are missed by the research community in their studies, and whether filter lists 
should still be used as the \emph{default tools} to detect trackers at scale.

{\bf Our contributions: }
To evaluate the effectiveness of filter lists, we propose a new, fine-grained behavior-based tracking detection. 
Our results are based on a stateful dataset of 8K domains with a total of 800K pages 
generating 4M third-party requests.
We make the following contributions:

{\em 1- We analyse all the requests and responses that lead to invisible pixels
(by ``invisible pixels'' we mean  $1\times1$  pixel images or images without content)}.
Pixels are routinely used by trackers 
to send information or third-party cookies 
back to their servers: the simplest way to do it is to create a URL containing useful information, 
and to dynamically add an image tag into a webpage. 
This 
makes invisible pixels \emph{the perfect suspects for tracking}
and propose a new 
classification of tracking behaviors.
Our results show that pixels are still widely deployed: 
they are present on more than 94\% of domains and constitute 35.66\% of all third-party images. 
We found out that pixels are responsible only for 23.34\% of tracking requests, and the most 
popular tracking content are scripts: a mere loading of scripts is responsible for 34.36\% of tracking requests.

{\em 2- We uncover hidden collaborations between third parties.}
We applied our classification 
on more than 4M third-party requests collected in our crawl. We have  detected new categories of tracking 
and collaborations between domains. 
We show that domains 
sync first party cookies through a \emph{first to third party cookie syncing}.
This tracking appears on 67.96\% of websites. 

{\em 3- We show that filter lists miss a significant number of cookie-based tracking. }
Our evaluation of the effectiveness of EasyList\&EasyPrivacy and Disconnect lists 
shows that they respectively miss 25.22\% and 30.34\% of the trackers that we detect.
Moreover, we find that if we combine all three lists, 379,245 requests 
originating from 8,744 domains still track users on 68.70\% of websites.

{\em 4- We show that privacy browser extensions miss a significant number of cookie-based tracking. }
By evaluating the popular privacy protection extensions: Adblock, Ghostery, Disconnect, and Privacy Badger, we show that Ghostery is the most efficient among them and that all extensions fail to block at least 24\% of tracking requests.

%% file: methodology.tex
\section{Methodology}
\label{sec:methodology}

To track users, domains deploy different mechanisms 
that have different impacts on the user's privacy.
While some domains are only interested in tracking the user within the same website,
others are recreating her browsing history by tracking her across sites.
In our study, by ``Web   tracking'' 
we refer to 
both within-site and 
cross-site tracking.\\
To detect Web  tracking, we first collect data from Alexa top 10,000 domains, then by analyzing the invisible pixels we define a new classification of Web   tracking behaviors
 that we apply
to the full dataset.
In this section, we explain the data collection process and the criteria
we used to detect identifier cookies and  cookie sharing.

\subsection{Data collection } 
\label{data}
\label{sec:crawling}

{\bf Two stateful crawls: }
We performed passive Web   measurements using the OpenWPM platform \cite{Engl-Nara-16-CCS}. 
It uses the Firefox browser, and 
provides browser automation by converting high-level commands into automated browser actions. 
We launched 
\emph{two stateful crawls on two different machines with different IP addresses}.
For each crawl, we used one browser instance 
and saved the state of the browser between websites.
In fact, measurement of Web  tracking techniques such as cookie syncing is 
based on re-using cookies stored in the browser, and hence 
it is captured more precisely in a stateful crawl.
\textbf{Full dataset: }
We performed a stateful crawl of  Alexa  top $10,000$ domains in February 2019 in France \cite{Alexa} from two different machines. 
Due to the dynamic behavior of the websites,
the content of a same page might differs every time this page is visited.
To reduce the impact of this dynamic behavior and reduce the difference between the two crawls,
we launched the two crawls at the same time.
For each domain, we visited the home page and the first $10$ links 
from the same domain.
The timeout for loading a homepage is set up to $90$s,  and the timeout for loading a link on the homepage is set up to $60$s.
Out of $10,000$ Alexa top domains, we successfully crawled $\totaldomainssucc$ domains with a total of $\totalpagessucc$ pages.

For every page we crawl, we store  the HTTP request (URL, method, header, date, and time), the HTTP response (URL,  method,  status code,  header, date,  and time), and the cookies (both set/sent and a copy of the browser cookie storage) to be able to capture the communication between the client and the server. We also store the body of the HTTP response if it's an image with a   \emph{content-length} less than 100 KB.
We made this choice to save storage space.
Moreover, in addition to HTTP requests, responses and cookies, we were only interested in the storage of invisible pixels. 
In our first dataset, named \emph{full dataset}, we capture all HTTP requests, responses, and cookies.

\textbf{Prevalence of invisible pixels: }
As a result of our crawl of \totalpagessucc\ pages, we have collected $\totalimages$ images detected using the field  \emph{content-type} in the HTTP header. We only stored images  with a \emph{content-length} less than $100$ KB.
These images represent  $89.83\%$ of the total number of delivered images.
Even though we didn't store all the images,
we were able to get the total number of delivered images
using the content-type HTTP header extracted from the stored HTTP responses.

\begin{figure}[th!]
\SHORTEN
 \centering
  \includegraphics[width=0.35\textwidth]{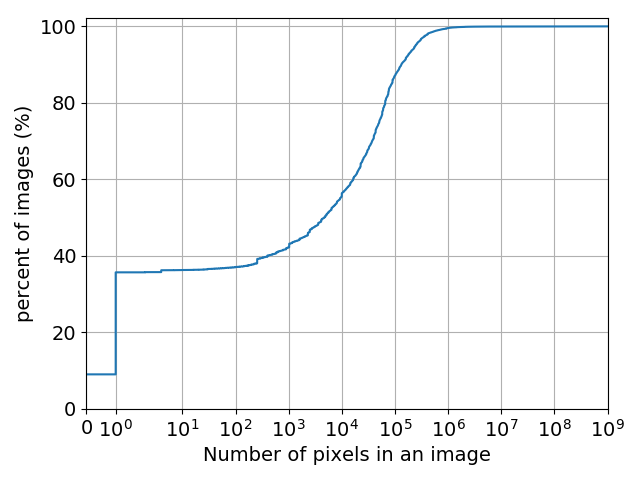}
\caption[Number of pixels in images]{Cumulative function of the number of pixels in images  with a  \emph{content-length} less than $100$ KB. {\normalfont \emph{$35.66$\% of the images are invisible pixels, $9.00\%$ 
have no content (they are shown as zero-pixel images),
and  $26.66\%$ 
are of size $1\times 1$ pixel.}}}
\label{fig:total_pixel}
\end{figure}

Figure \ref{fig:total_pixel} shows the distribution of the number of pixels in all collected images. 
We notice that invisible pixels ($1\times 1$ pixels and images with no content) represent $35.66$\% of the total number of collected images.

We found that out of $\totaldomainssucc$ successfully crawled domains, $8,264$ ($94.51$\%) domains contain at 
least one page with one invisible pixel. 
By analyzing webpages independently, we found that  
$92.85$\% out of $\totalpagessucc$ visited pages include at least one invisible pixel.

\textbf{Invisible pixels subdataset:}
The invisible pixels do not add any content to the Web   pages. However, they are widely used in the Web. They generally allow the third party to send some information using the requests sent to retrieve the images.
Moreover,  the user is unaware of their existence.
Hence, every invisible pixel represents a threat to the user privacy.
We consider the set of requests and  responses used to serve the invisible pixels as a ground-truth dataset that we call 
\emph{invisible pixels dataset}. The study of this \emph{invisible pixels dataset} allow us to excavate the tracking behaviors of third party domains in the web.

\subsection{Detecting identifier cookies } 
\label{DetectingcookieID}

Cookies are a classical way to track users in the Web. 
A key task to detect this kind of 
tracking is to be able to detect cookies used to store identifiers.
We will refer to these cookies as \emph{\id}s.
In order to detect {\id}s, we analyzed data extracted from the two simultaneous crawls performed from two different machines.
We refer to the owner of the cookie as host, 
and we define a cookie instance as (host, key, value).

\textbf{We compare cookies instances between the two crawls: }
A tracker associates different identifiers to different users in order to distinguish them.
Hence, an {\id} should be unique per user (user specific). 
We analyzed the \totaldomainssucc\ crawled websites where we have a total of $607,048$ cookies instances
belonging to $179,580$ (host, key) pairs. 
If an identical cookie instance appears in the two crawls, 
that is, the host, key and value of both cookies are identical,
we consider that the cookie is not used for tracking.
We refer to such cookies as  \emph{safe cookies}.
We extracted $108,252$ safe cookies from our dataset. 
They represent 17.83\% of the total number of cookies instances.

Due to the dynamic behavior of websites,
not all cookies appear in both crawls.
We mark the
cookies (host, name) that appear only in one crawl as unknown cookies.
In total, we found $15,386$ unknown cookies (8.56\%).
We exclude these cookies from our study.

\textbf{We don't consider the cookie lifetime: } 
The lifetime of the cookie  
is used  to detect {\id}s
 in related works \cite{Acar-etal-14-CCS,Engl-etal-15-WWW,Engl-Nara-16-CCS}.
Only cookies that expire at least a month after being placed are considered as {\id}s.  
In our study,  we 
don't put any boundary on the cookie lifetime 
because domains can continuously update 
cookies with a short lifetime and do the mapping 
of these cookies on the server side which will allow a long term tracking.

\textbf{Detection of cookies with {\id} as key: }
We found that some domains store the \id\ as part of the cookie key.
To detect this behavior,   
we analyzed the  cookies with the same host and value and  different keys across the two crawls.  
We found $5,295$ (0.87\%) cookies instances  with {\id} as key.
This behavior was performed by $966$ different domains.
Table \ref{tab:host} in Appendix presents the top $10$ domains involved.
The cookies with  {\id} as key  represent only  0.87\% of the total number of cookies.
Therefore, we will exclude them from our study.

\subsection{Detecting identifier sharing} 
\label{DetectingID}

Third party trackers not only collect data about the users, but also exchange this data 
to build richer users profiles.
Cookie syncing is a common technique used to exchange user identifiers stored in cookies.
To detect such behaviors, we need to detect the \emph{identifier cookies} shared between domains.
A cookie set by one domain cannot be accessed by another domain
because of the cookie access control and Same Origin Policy  
\cite{SameOriginPolicy}.
Therefore, trackers need to pass identifiers 
through the URL parameters.

\begin{figure}[t]
\centering
\includegraphics[width=0.35\textwidth]{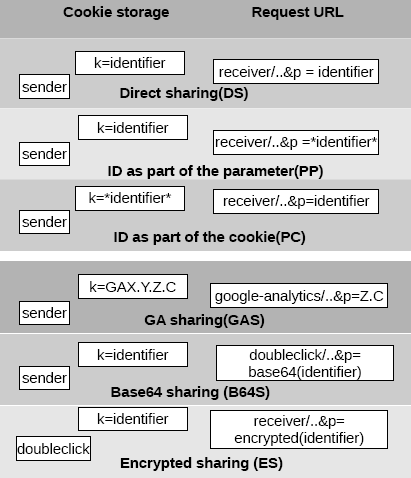}
\caption[lenght]{
Detecting identifier sharing. ''sender'' is the domain that  owns  the cookie and  triggers the request, ''receiver'' is the domain that receives this request, and ''identifier'' is the \emph{identifier cookie} value.  ''*'' represents any string. }
\label{fig:cookiesharing}
\SHORTEN
\end{figure}

Identifier sharing can be done in different ways: 
it 
can be sent in clear as a URL parameter value, 
or 
in a specific format, encoded or even encrypted.
To detect identifiers,
we take inspiration from 
 \cite{Engl-Nara-16-CCS,Acar-etal-14-CCS}.
 We  split  
cookies and 
URL parameter values using as delimiters 
 any character not 
in [a-zA-Z0-9,'-','\_','.'].
Figure \ref{fig:cookiesharing} shows 
six different techniques we deployed  to detect 
identifier sharing. 
The first three methods 
are generic: either the 
identifier is sent as the parameter value, 
as part of the parameter value 
or it's stored as part of the cookie value and sent as parameter value.

We noticed that the requests for invisible images, 
where we still didn't detect any cookie sharing, originate mostly from 
\domain{google-analytics.com} and \domain{doubleclick.net}. 
Indeed, 
these domains are prevalent in 
serving invisible pixels across websites (see Figure \ref{fig:resp_dom} in Appendix).
We therefore base the next techniques on these two use cases.
First, we notice that first party cookies set by \domain{google-analytics.com} have the format GAX.Y.Z.C, 
but the identifier  sent to it are of the form Z.C. We therefore detect this particular type of cookies, 
that were not detected in previous works that rely on delimiters  (\textbf{GA sharing}). 
Second, by base64 decoding the value of the parameter sent to \domain{doubleclick.net}, we detect the encoded sharing(\textbf{Base64 sharing}). 
Finally, by relying on 
Doubleclick documentation \cite{Doubleclick} we infer that encrypted cookie was shared(\textbf{Encrypted sharing}).
For more details see the Section \ref{sharingapp} in  the Appendix.

\subsection{Limitations}
We detected six different techniques used to share the {\id}.
However, trackers 
may 
encrypt the cookie before sharing it.
In this work, we only detected encrypted cookies when it's shared following a 
specific semantic set by \domain{doubleclick}~\cite{Doubleclick}.

We do not
inspect the payload of POST requests
that could be used to share the {\id}.
For example, it's known that \domain{google-analytics.com} sends the {\id} as part of the URL parameters
with GET requests or in the payload of the POST requests \cite{analyticspost} -- 
we 
do not detect such a case in this work.

To detect the sender of the request in case of inclusion, 
we use the referer field.
Therefore, we may miss to interpret
who is the effective initiator of the request, 
it can be either the first party or an included script.

%% file: results.tex
\section{Overview of tracking behaviors} 
\label{sec:results}

In Section~\ref{sec:crawling}, we detected that invisible pixels are widely present 
on the Web and are perfect suspects for tracking. 
In this Section, we 
detect the different tracking behaviors by analyzing the 
\emph{invisible pixels dataset}.

In total, we have $747,816$ third party requests leading to invisible images. By analyzing these 
requests, we detected $6$ categories of tracking behaviors  in $636,053$ (85.05\%) 
requests that lead to invisible images.

We further group these categories 
into three main classes: 
explicit cross-site tracking (Section \ref{explicit}), cookie syncing (Section \ref{syncing}), 
and analytics  (Section \ref{analytics}). 
In the following, we call \textit{{\ourmethod} } our detection method of these behaviors.

After defining our classification using the \textit{invisible pixels dataset}, 
we apply it
on the \textit{full dataset} where we have a total of 4,216,454 third-party requests 
collected from  \totalpagessucc\ pages on the   $\totaldomainssucc$\ domains successfully crawled.
By analyzing these requests, we detected  6 tracking behaviors in $2,724,020$ ($64.60$\%)
requests.

\begin{figure}[!t]
\centering
\includegraphics[width=0.5\textwidth]{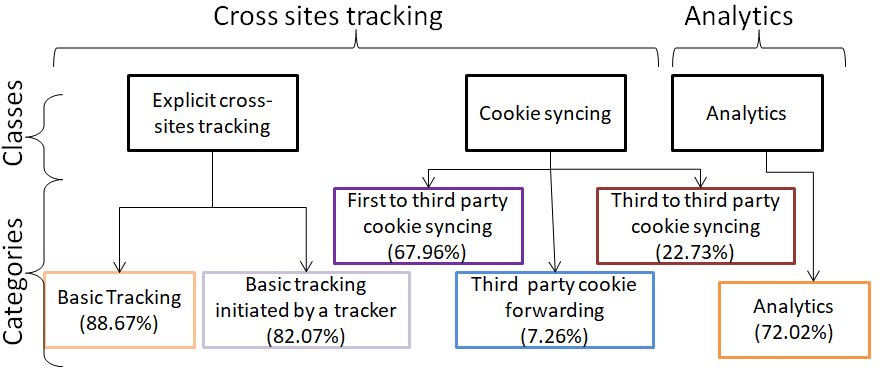}
\caption{\label{fig:categorizationoverview}Classification overview. {\normalfont \emph{(\%) represents the percentage of domains out of \totaldomainssucc\  where we detected the tracking behavior. 
A tracking behavior is performed in a domain if 
it's detected in at least one of its pages.}} }
\end{figure}

Figure \ref{fig:categorizationoverview} presents an overview of all classes (black boxes) and 
categories 
of tracking behaviors 
and their prevalence in the full dataset. 
%
Out of  $\totaldomainssucc$\  crawled domains, we identified at least one form of tracking in  91.92\% domains. 
We  further analyzed prevalence of each tracking category that we report in Section~\ref{sec:categorization}. 
We found out that \emph{first to third party cookie syncing (see Sec.~\ref{sec:firsttothird})
appears on 
67.96\% of the domains!}

\begin{figure}[htp]
\centering

\subfloat{%
  \includegraphics[width=0.4\textwidth]{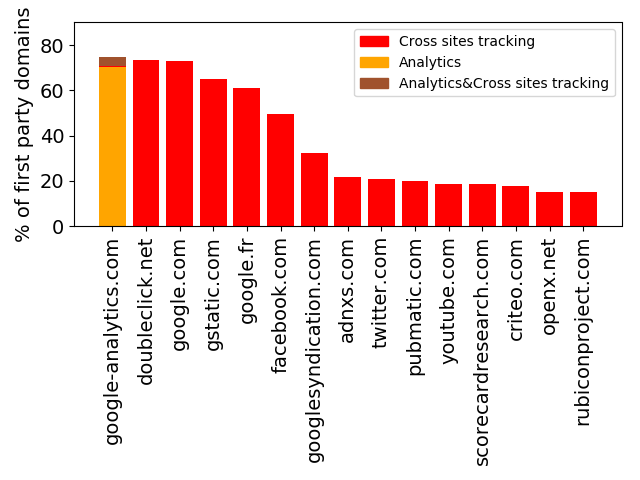}%
}
\vspace{-0.2 cm}
\subfloat{%
  \includegraphics[width=0.4\textwidth]{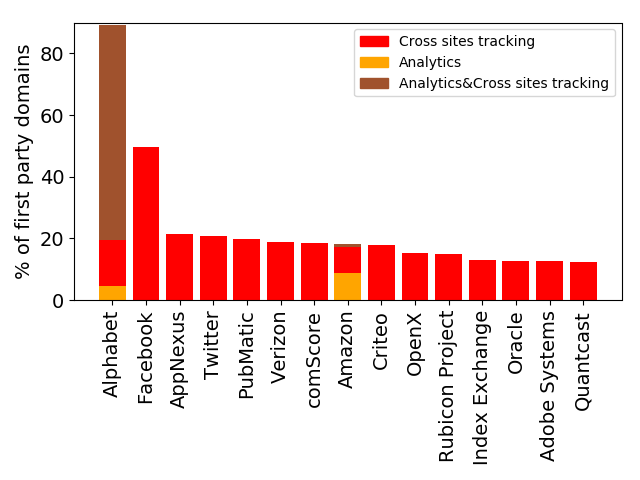}%
}
\caption{\label{fig:inclusion}
Top 15 domains and companies involved in analytics, cross-site tracking, or both on the same first-party domain.
}
\end{figure}

In addition, we analyzed the most prevalent domains involved in either cross-site tracking, 
analytics, or both behaviors. 
Figure~\ref{fig:inclusion} demonstrates 
that a third party domain may have 
several behaviors. 
For example, we detect that \domain{google-analytics.com}  
exhibits both cross-site tracking and analytics behavior. 
This variance of behaviors is due to  the web site developer,
as it's the case for cookie syncing and analytics behaviors.
It can also be due to  the domain's partners as it's the case for cookie forwarding.
Google-analytics in that case is included by another third party, the developer is not necessarily aware of this practice.

\begin{table}[!h]
    \centering
    \begin{tabular}{l|c}
         \textbf{Content type}& \textbf{\% requests} \\ \hline
         Script &  34.36\%  \\ \hline
        Invisible images &  23.34 \% \\ \hline
        Text/html & 20.01\%  \\ \hline 
        Big images & 8.54 \% \\ \hline
        Application/json &  4.32\%  \\ \hline
    \end{tabular}
    \caption{Top 5 types of content used in the $2,724,020$  third party tracking requests. }
    \label{tab:types}
\SHORTEN
\end{table}{}

We found that not all the tracking detected in the \emph{full dataset} is based on invisible pixels. 
We extracted the type of the content served by the tracking requests 
using the HTTP header \emph{Content-Type}.
Table \ref{tab:types} presents the top 5 types of content used for tracking.
Out of the $2,724,020$ requests involved in at least one  tracking behavior 
in the full dataset, the top content delivered by tracking requests 
is scripts ($34.36$\%), while the second most common content 
is invisible pixels 
($23.34$\%).
We also detected other content used for tracking purposes such as visible images.

\section{Classification of tracking }
\label{sec:categorization}

In this Section, we explain all the categories of tracking 
behaviors presented in Figure \ref{fig:categorizationoverview} that we have 
uncovered by studying the \emph{invisible pixels} dataset. 
For each category, we start by explaining the tracking behavior, 
we then give its privacy impact on the user's privacy, 
and finally we present the results from the \emph{full dataset}.

\subsection{Explicit cross-site tracking }
\label{explicit}

Explicit cross-site tracking class includes two categories:
\textit{basic tracking} and \textit{basic tracking initiated by a tracker}. 
In both categories, we do not detect 
cookie syncing 
that we analyze separately in Section \ref{syncing}.

\subsubsection{Basic tracking } \label{bt}

\textit{Basic tracking} is the most common tracking category 
as we see from Figure \ref{fig:categorizationoverview}.

{\bf Tracking behavior: }
Basic tracking happens when a  third party domain, say \domain{A.com}, 
sets an \id\ in the user's browser. 
Upon  a visit to a webpage with content from  \domain{A.com}, 
a request is sent to  \domain{A.com} with its  cookie. 
Using this cookie, \domain{A.com} identifies the user across all websites 
that include content from \domain{A.com}.

\textbf{Privacy impact:} \textit{Basic tracking} is the best known tracking technique that
allows third parties to track the user across websites, 
hence to recreate her browsing history.  
However, third parties are able to track the user 
only on the websites where their content is  present.

\begin{figure}[htp]

\centering

\subfloat{%
\includegraphics[width=0.37\textwidth]{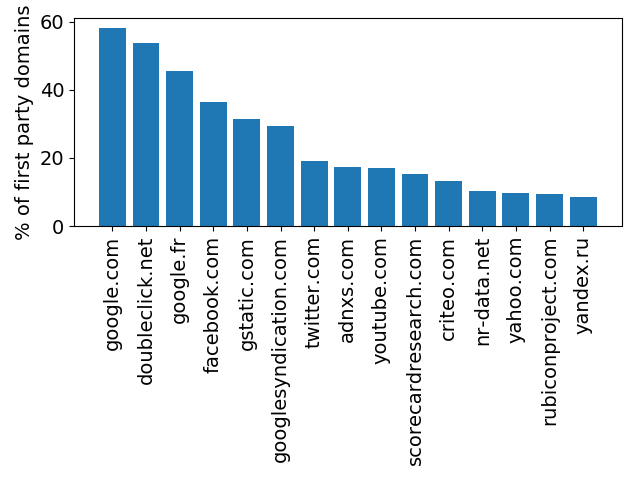}%
}
\vspace{-0.2 cm}
\subfloat{%
 \includegraphics[width=0.37\textwidth]{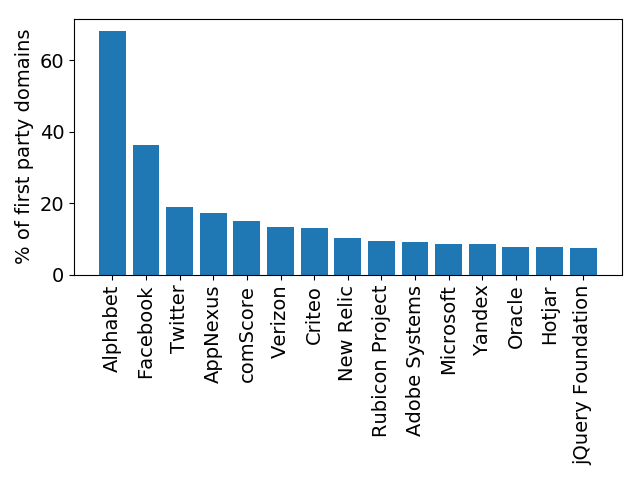}%
}
\caption{\label{fig:crosstracker} Basic tracking: Top 15 cross-site trackers and companies in charge of the trackers included in $\totaldomainssucc$ domains.}

\end{figure}

\textbf{Results: }We detected basic tracking in 88.67\% of visited domains. 
In total, we found $5,421$ 
distinct third parties performing basic tracking. 
Figure \ref{fig:crosstracker} shows the top domains involved in basic tracking. 
We found that  \domain{google.com} alone is tracking the user on over $5,079$ (58.08\%) domains.  
This percentage becomes more important if we consider the company instead of the domain (Figure \ref{fig:crosstracker}).
By considering companies instead of domains, 
we found that, by only using the \textit{basic tracking} \domain{Alphabet} (the owner of \domain{Google})  
is tracking users in 68.30\% of Alexa top 8K websites.

\subsubsection{Basic tracking initiated by a tracker} \label{btredi}
When the user visits 
a website that includes content from a third party, the third party 
can redirect the request to a second third party tracker or include it.
The second  tracker  will associate his own {\id} to the user. 
In this case the second tracker is not directly embedded by the first party and yet it can track her.

{\bf Tracking behavior: }
\textit{Basic tracking initiated by a tracker} happens when a basic tracker 
is included in a website by another basic tracker.

\textbf{Privacy impact:} By  redirecting to each other, trackers trace the user activity on a larger number of websites. They gather the browsing history of the user on websites where at least one of them is included. The impact of these behaviors on the user's privacy could be similar to  the impact of cookie syncing. 
In fact, by mutually including each other on websites, each tracker can recreate the combination of what both partners have collected using basic tracking.
Consequently, through  \textit{basic tracking initiated by a tracker}, 
trackers get to know the website visited by the users,
without being included in it
as long as this website includes one of the tracker's partners.
Hence, through this tracking technique, 
the user's browsing history is shared instantly without syncing cookies.

\begin{figure}[htp]

\centering

\subfloat{%
\includegraphics[width=0.37\textwidth]{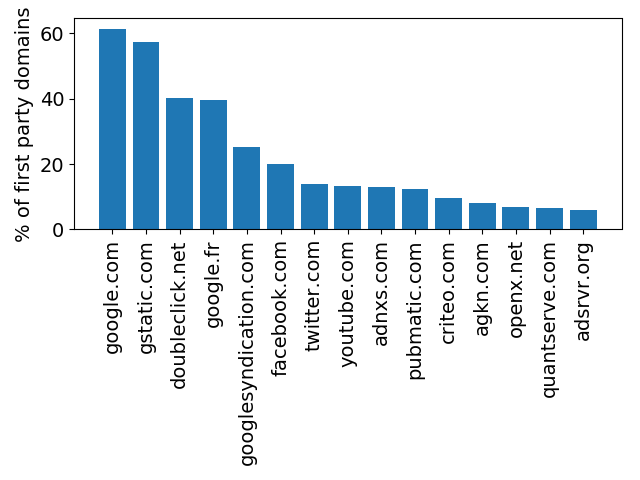}%
}
\vspace{-0.2 cm}
\subfloat{%
\includegraphics[width=0.37\textwidth]{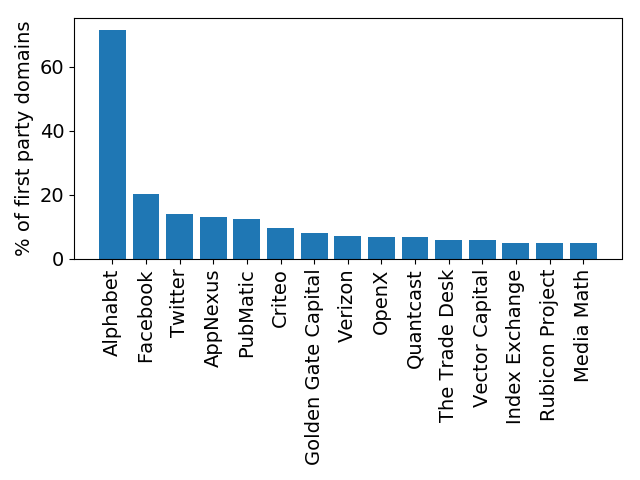}%
}
\caption{ \label{fig:inclutracking}Basic tracking initiated by a tracker: Top 15 trackers and companies included in $\totaldomainssucc$ domains.}

\end{figure}

\begin{table}[h!]
\centering
\begin{tabular}{l|c}
\hline
Partners & \#  requests \\ \hline

pubmatic.com $\leftrightarrow$ doubleclick.net  &  4,392  \\ \hline
criteo.com $\leftrightarrow$ doubleclick.net  &  2,258  \\ \hline
googlesyndication.com $\leftrightarrow$ adnxs.com  &  1,508  \\ \hline
googlesyndication.com $\leftrightarrow$ openx.net  &  1,344  \\ \hline
adnxs.com $\leftrightarrow$ doubleclick.net  &  1,199  \\ \hline
rubiconproject.com $\leftrightarrow$ googlesyndication.com  &  1,199  \\ \hline
doubleclick.net $\leftrightarrow$ yastatic.net  &  979  \\ \hline
doubleclick.net $\leftrightarrow$ demdex.net  &  790  \\ \hline
adnxs.com $\leftrightarrow$ amazon-adsystem.com  &  760  \\ \hline
rfihub.com $\leftrightarrow$ doubleclick.net  &  685  \\ \hline

\end{tabular}
\caption{\label{tab:dc}Basic tracking initiated by a tracker: Top $10$ pairs of partners from different companies that include each other.
($\leftrightarrow$) both ways inclusion.}
\SHORTEN
\end{table}

\textbf{Results:} We detected Basic tracking initiated by a tracker in $82.07$\% of the domains. 
From Figure \ref{fig:inclutracking}, we can notice that \domain{google.com} 
is the top tracker included by other third parties. 
By only relying on its partners, without being directly included by the developer,
\domain{google.com} is included in over $5,374$ (61.45\%) of the Alexa top 8k domains and its owner company \domain{Alphabet} is included in over $71.56\%$ of the visited domains.
\domain{Google.com} is included by $295$ different third party trackers in our dataset. 
In our results, we found that \domain{doubleclick.net} and \domain{googlesyndication.com}, both owned by Google, are the top domains 
including each other (176,295 requests in our dataset).
Table \ref{tab:dc}  presents the top $10$ pairs of partners from different companies that are mutually including each other on websites.
Note that in Table \ref{tab:dc} we don't report mutual inclusion of domains that belong to the same company.

\subsection{Cookie syncing} \label{syncing} 
To create a more complete profile of the user,
third party domains need to  merge profiles they collected on different websites.
One of the most known techniques to do so is cookie syncing.
We separate the previously known technique of cookie syncing \cite{Acar-etal-14-CCS, Engl-Nara-16-CCS} into two distinct categories, 
\textit{third to third party cookie syncing} and \textit{third party cookie forwarding}, 
because of their different privacy impact.
We additionally detect a new type of cookie syncing that we call \textit{first to third party cookies syncing}.

\subsubsection{Third to third party cookie syncing}

When two third parties have an {\id} in a user's browser 
and need to merge user profile, they use third to third party cookie syncing.

\begin{figure}[th!]
\centering
\includegraphics[width=0.4\textwidth]{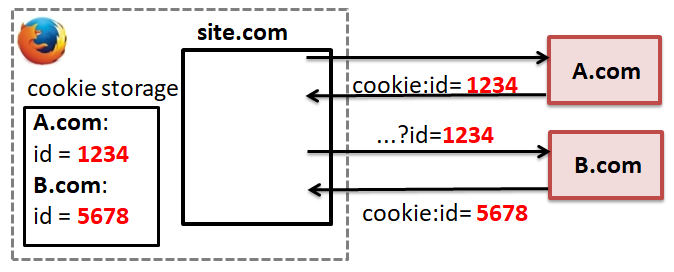}
\caption{\label{fig:syncing} Third to third party cookie syncing behavior. 
}
\SHORTEN
\end{figure}

\textbf{Tracking explanation: }Figure \ref{fig:syncing} demonstrates cookie syncing\footnote{Notice that in figures that explain the tracking behaviors, we show cookies 
only in the response, and never in a request. This actually represents both cases 
when cookies are sent in the request and also set in the response. }. The first party domain includes a content having as source the first third party \domain{A.com}. 
A request is then sent to \domain{A.com} to fetch the content.
Instead of sending the content, \domain{A.com} decides to redirect to \domain{B.com} and in the redirection request sent to \domain{B.com}, \domain{A.com} includes the identifier it associated to the user.
 In our example, \domain{B.com} will receive the request \url{B.com?id=1234}, where \textit{1234} is the identifier associated by \domain{A.com} to the user. Along with the request, \domain{B.com} will receive its cookie \textit{id = 5678}, which will allow \domain{B.com} to link the two identifiers to the same user.

\textbf{Privacy impact: } \textit{Third to third party cookie syncing} is one of the most harmful tracking techniques that impact the user's privacy. In fact, third party cookie syncing can be seen as a set of trackers performing \textit{basic tracking} and then exchanging the data they collected about the user. 
It's true that a cross sites tracker recreates part of the user's browsing history but this is only possible on the websites on which it was embedded. Using cookie syncing, a tracker does not only log the user's visit to the websites where it's included, but it can also log her visits to the websites where its partners are included.
What makes this practice even more harmful is when a third party has several partners with whom it syncs cookies. 
One example of such behavior is  \domain{rubiconproject.com}, that syncs its {\id}  with 7 partners: \domain{tapad.com}, \domain{openx.net}, \domain{imrworldwide.com},  \domain{spotxchange.com}, \domain{casalemedia.com}, \domain{pubmatic.com} and \domain{bidswitch.net}.

\begin{table}[h!]
\centering
\begin{tabular}{p{4.1cm}|p{1cm}|p{2cm}}
\hline

Partners & \#  requests &  Sharing technique \\ \hline
adnxs.com $\rightarrow$ criteo.com & 1,962 & $\rightarrow$DS \\ \hline
doubleclick.net $\rightarrow$  facebook.com & 789 & $\rightarrow$DS \\ \hline
casalemedia.com $\rightarrow$ adsrvr.org & 778 & $\rightarrow$DS \\ \hline
mathtag.com$\leftrightarrow$adnxs.com & 453 & $\rightarrow$DS \\ \hline
pubmatic.com $\rightarrow$ lijit.com & 321 & $\rightarrow$DS \\ \hline
adobedtm.com  $\rightarrow$ facebook.com & 269 & $\rightarrow$DS \\ \hline
doubleclick.net $\leftrightarrow$ criteo.com & 250 & $\rightarrow$ DC, PCS; \hspace{0.2 cm} $\leftarrow$ DS \\ \hline
mmstat.com $\rightarrow$ cnzz.com & 233 & 
 $\rightarrow$ DC \\ \hline

sharethis.com $\rightarrow$ agkn.com &  233 & 
 $\rightarrow$ DC \\ \hline

mathtag.com $\rightarrow$  lijit.com & 109 & 
 $\rightarrow$ DC  \\ \hline
\end{tabular}
\caption{\label{tab:syncing}Third to third party cookie syncing: Top 10 partners. The arrows represent the flow of the cookie synchronization, ($\rightarrow$) one way matching or ($\leftrightarrow$) both ways matching. DS (Direct sharing), PCS (ID as part of the cookie), PPS (ID sent as part of the parameter) are different sharing techniques described in  Figure \ref{fig:cookiesharing}.}
\SHORTEN
\end{table}

\textbf{Results: }We detected third to third party cookie syncing in 22.73\%  websites.
We present in Table \ref{tab:syncing} the top $10$ partners that we detect as performing cookie syncing.
In total, we have detected $1,263$ unique partners performing cookie syncing. The syncing could be done in both ways, as it's the case for  \domain{doubleclick.net} and \domain{criteo.com}, or in one way, as it's the case for \domain{adnxs.com} and \domain{criteo.com}. In case of two ways matching, we noticed that the two partners can perform different identifier sharing techniques. 
 We see the complexity of the third to third party cookie syncing that involves a large variety of sharing techniques.

\SHORTEN
\subsubsection{Third-party cookie forwarding}

The purpose of the collaboration between third party domains in \textit{third party cookie forwarding} is to instantly share the browsing history.
Cookie forwarding has always been  called “syncing” while 
instead it simply enables a third party to reuse an identifier of a tracker, without actually syncing its own identifier.

\begin{figure}[th!]
\centering
\includegraphics[width=0.4\textwidth]{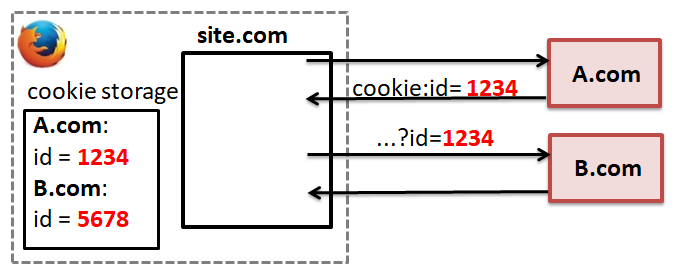}
\caption{\label{fig:ttfrwd}Third party cookie forwarding behavior. 
}
\SHORTEN
\end{figure}

\textbf{Tracking explanation: } 
The first party domain \domain{site.com} includes \domain{A.com}'s content.
To get the image, a request is sent to \domain{A.com} along with its cookie. \domain{A.com} then redirects the request to its partner (\domain{B.com}) and sends  the \id\ it associated to the user (\textit{1234}) as part of the URL parameters (Figure \ref{fig:ttfrwd}).

\textit{Third party cookie forwarding} differs from  \textit{Third to third party cookie syncing} depending on whether there  is a cookie set by the receiver  in the browser or not. 
This category is similar to third-party advertising networks in Roesner {\al}  and Lerner \al's works  \cite{Roes-etal-12-NSDI} \cite{Lern-etal-16-USENIX}, in the sense that we  have a collaboration of third-party advertisers.
However, in our study we check that the second tracker do not use its own cookie to identify the user. This means that this tracker (\domain{B.com}) is  relying on the first one (\domain{A.com}) to track the user. In fact,  \domain{B.com} uses \domain {A.com}'s identifier to recreate her browsing history.

\textbf{Privacy impact: } \textit{Third party cookie forwarding} allows trackers to instantly share the browsing history of the user. \domain{A.com} in Figure \ref{fig:ttfrwd} does not only associate an  \id\ to the user, but it also redirect and shares this \id\ with it's partner. This practice allows both \domain{A.com} and \domain{B.com} to track the user across websites. From a user privacy point of view, \textit{third party cookie forwarding} is not as harmful as cookie syncing, because the second tracker in this case does not contribute in the user's profile creation but  passively receives the user's browsing history from the first tracker.

\textbf{Results: }We detected third party cookie forwarding in 7.26\% of visited websites.
To our surprise, the top domain receiving {\id} from third parties is \domain{google-analytics.com} (Figure \ref{fig:receivers_transfer_cookie_no_fp} in Appendix). 
\domain{Google-analytics.com} is normally included by domains owners  to get analytics of their websites, it's known as a \emph{within domain tracker}. But in this case,  \domain{google-analytics.com} is used by the third party domains.  The third party is  forwarding its third party cookie to \domain{google-analytics.com} on different websites, consequently \domain{google-analytics.com} in this case is   tracking the user across websites.
This behavior was discovered by Roesner \emph{\al}  \cite{Roes-etal-12-NSDI}. They reported this behavior in only a few instances, but in our dataset we found $386$ unique partners that forward cookies, among which  $271$ are forwarding cookies to  \domain{google-analytics.com}. 
In Table \ref{tab:gafrwd} (Appendix), we present the top $10$ third parties forwarding cookies to google-analytics service.

\subsubsection{First to third party cookie syncing}
\label{sec:firsttothird}
 
In this category, we detect that first party cookie get synced with third party domain.
\begin{figure}[th!]
\centering
\includegraphics[width=0.4\textwidth]{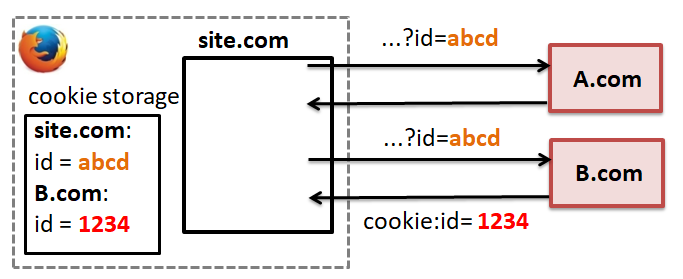}
\caption{\label{fig:sync_fp} First to third party cookie syncing behavior. 
}
\SHORTEN
\end{figure}

\textbf{Tracking explanation: } Figure \ref{fig:sync_fp} demonstrates the cookie syncing of the first party cookie. 
The first party domain \domain{site.com} includes a content from 
\domain{A.com?id=abcd}, where \domain{A.com} is a third party and \textit{abcd} is the first party \id\ of the user set for 
\domain{site.com}.
\domain{A.com} receives the first party cookie \emph{abcd} in the URL parameters,
and then redirects the request to \domain{B.com}.
As part of the request redirected to \domain{B.com},
\domain{A.com} includes the first party {\id}.
\domain{B.com} sets its own \id\ \textit{1234} in the user's browser. 
Using these two identifiers (the first party's identifier \textit{abcd} received in the URL parameters and its own identifier \textit{1234} sent in the cookie),  
\domain{B.com} can 
create a matching table
that allows \domain{B.com} to link both 
identifiers 
to the same user.

The first party cookie can also be shared directly by the first party service (imagine Figure \ref{fig:sync_fp} where \domain{A.com} is absent).
In that case, \domain{site.com} includes content from \domain{B.com}
and as part of the request sent to \domain{B.com}, 
\domain{site.com} sends the first party {\id} \textit{1234}.
\domain{B.com} sets its own \id\ \textit{1234} in the user's browser.
\domain{B.com} can now link the two identifiers to the same user.

\textbf{Privacy impact: }In our study, we differentiate the case when 
cookie shared is a first party cookie and when it is a third party cookie.
We made this distinction because, the kind and the sensitivity of the data shared differs in the two cases. 
Using this tracking technique,
first party websites get to sync cookies with third parties.
Moreover, pure analytic services allow to sync in-site history with cross-site history.

\begin{table}[h!]
\centering
\begin{tabular}{l c}
\hline
Partners & \#  requests  \\ \hline
 \multicolumn{2}{l}{\bf First party cookie synced  through an
 } 
 \\
  \multicolumn{2}{l}{\bf  
 \hspace{1.5 cm} intermediate service} 
\\ \hline
google-analytics.com $\rightarrow$ doubleclick.net & 8,297   \\  \hline  
 \multicolumn{2}{l}{\bf Direct First to third party cookie syncing } 
\\ \hline
hibapress.com $\rightarrow$ criteo.com & 460 \\ 
alleng.org $\rightarrow$ yandex.ru & 332 \\
arstechnica.com $\rightarrow$ condenastdigital.com & 243 \\ 
thewindowsclub.com $\rightarrow$ doubleclick.net & 228 \\ 
digit.in $\rightarrow$ doubleclick.net & 224 \\
misionesonline.net $\rightarrow$ doubleclick.net & 221 \\ 
wired.com $\rightarrow$ condenastdigital.com & 219 \\ 
newyorker.com $\rightarrow$ condenastdigital.com & 218 \\ 
uol.com.br $\rightarrow$ tailtarget.com & 198 \\

\end{tabular}
\caption{\label{tab:ftps} First to third party cookie syncing: Top $10$ partners.}
\vspace{-0.7cm}
\end{table}

\textbf{Results: }We detected 
first  to third party cookie syncing in 67.96\% of visited domains.
In Table \ref{tab:ftps}, we present the top $10$ partners syncing first party cookies. 
We differentiate the two cases: 
(1) 
first party cookie synced through an intermediate service (as shown in Figure \ref{fig:sync_fp})
and (2) 
first party cookie synced directly from the  first party domain. 
In total we found $17,415$ different partners involved. 
The top partners are \domain{google-analytics.com} and \domain{doubleclick.net}. 
We found that \domain{google-analytics.com} first receives the cookie
as part of the URL parameters.
Then,
through a redirection process, \domain{google-analytics.com} transfers the first party cookie to \domain{doubleclick.net} that inserts or receives an \id\ in the user's browser.
We found out that \domain{google-analytics.com} is 
triggering such first party cookie syncing on $38.91$\% of visited websites.

\subsection{Analytics category} \label{analytics}
Instead of measuring website audience themselves, websites today use third party analytics services. Such services provide reports of the website traffic by tracking the number of visits, the number of visited pages in the website, etc.
The first party website includes content from the third party service on the pages it wishes to analyze the traffic.

\begin{figure}[th!]
\centering
\includegraphics[width=0.4\textwidth]{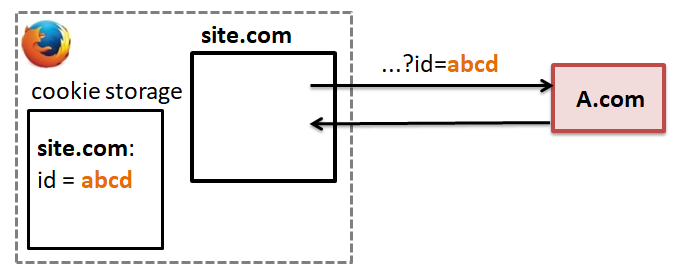}
\caption{\label{fig:analytics1} Analytics behavior. 
}
\SHORTEN
\end{figure}

\textbf{Tracking explanation: }Figure \ref{fig:analytics1} shows the analytics category where the domain directly visited by the user (\domain{site.com}) owns a cookie containing a unique identifier in the user's browser.  Such cookie is called a first party {\id}. 
This cookie is used by the third party (\domain{A.com}) to uniquely identify the visitors within \domain{site.com}.
The first party website makes a request to the third party to get the content and uses this request to share the first party \id.

\textbf{Privacy impact: }In analytic behavior, the third party domain is not able to track the user across websites because it does not set its own cookie in the user's browser. Consequently, for this third party, the same user will have different identifiers in different websites. 
However, using the first party \id\ shared by the first party, the third party can identify the user within the same website.
From a user point of view, analytics behavior is not as harmful as the other tracking methods. The analytics service can not recreate the user's browsing history but it can only track her activity within the same domain, which could be really useful for the website developer.

\textbf{Results: }We detected analytics in 72.02\% of the visited domains. 
We detect that \domain{google-analytics.com} is the most common analytics service.
It's used on $69.25$\% of the websites.
The next most popular analytics is \domain{alexametrics.com}, 
it's prevalent on $9.10$\% of the websites (see Figure \ref{fig:recanaly} in the Appendix).

%% file: comparing.tex

\section{Are filter lists effective at detecting trackers?}

 \label{sec:comparing}

Most of the state-of-the-art works that aim at measuring 
trackers at large scale rely on \emph{filter lists}. 
In particular, EasyList~\cite{easylist},  EasyPrivacy~\cite{easyprivacy} 
and \DISC~\cite{DisconnectList} lists became the \emph{de facto} approach to detect third-party tracking requests 
in the privacy and measurement communities 
\cite{Engl-Nara-16-CCS,
Bash-etal-16-USENIX,Laui-etal-17-NDSS,Raza-etal-18-NDSS,Ikra-etal-17-PETS,
Engl-etal-18-PETS,Bash-Wils-18-PETS,Bash-etal-18-IMC,Iord-etal-18-IMC}. 
Nevertheless, 
filter lists detect only known tracking and ad-related requests, 
therefore 
a tracker can easily 
 avoid 
 this detection 
by registering 
a new domain. 
Third parties can also incorporate tracking behavior into functional website content, 
which could not be blocked by filter lists because blocking functional content would harm user experience.
Therefore, it is interesting to evaluate how effective are filter lists at detecting trackers, how 
many trackers are missed by the research community in their studies, and whether filter lists 
should still be used as the \emph{default tools} to detect trackers at scale.

In this Section, we analyze how 
effective are filter lists at detecting third-party trackers. 
Contrary to Merzdovnik {\etal}'s work \cite{Merz-etal-17-EuroSnP}, 
which  measured blocking of third party requests without identifying whether such requests are tracking or not,
we compare 
all the cross-site tracking and analytics behavior reported in Section~\ref{sec:categorization} 
(that we unite under one detection method, that we call \ourmethod)
with the third-party trackers detected by filter lists. 
{EasyList and EasyPrivacy (\ELEP)} and {\disc} filter lists 
in our comparison were extracted in April 2019.
We use the Python library \emph{adblockparser} \cite{adblockparser}, to determine 
if a request would have been blocked by \ELEP. For \disc\, we compare to the domain name 
of the requests 
(the \disc\ list contains full domain names, while \ELEP\ are lists 
of regular expressions that require parsing). 

For the  comparison, we used the \emph{full dataset} of 
$4,216,454$ third party requests collected from 
$\totalpagessucc$ pages of 
 {$\totaldomainssucc$}  successfully crawled domains. 

\begin{figure}[th!]
\SHORTEN
\subfloat[
EasyList and EasyPrivacy\label{fig:compareELEP}]{\includegraphics[width=0.25\textwidth]{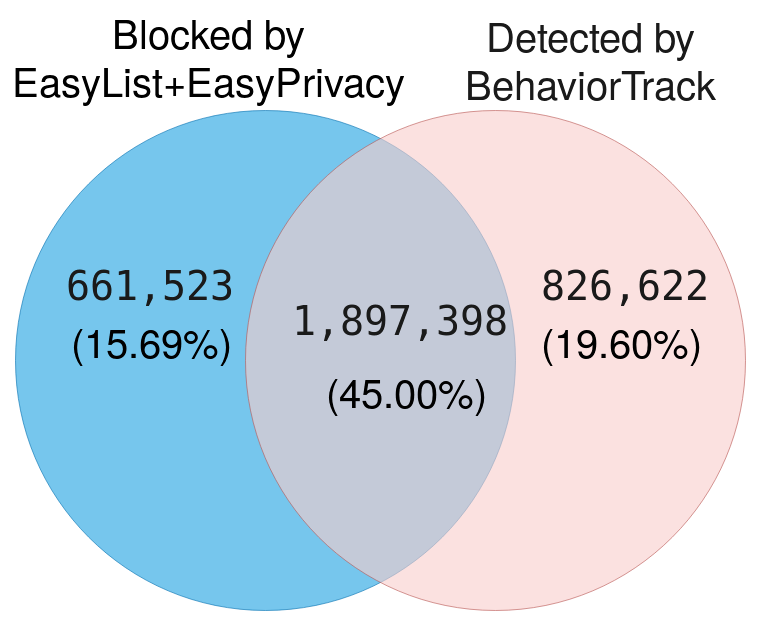}} 
\subfloat[
Disconnect\label{fig:compareDISC}]{\includegraphics[width=0.25\textwidth]{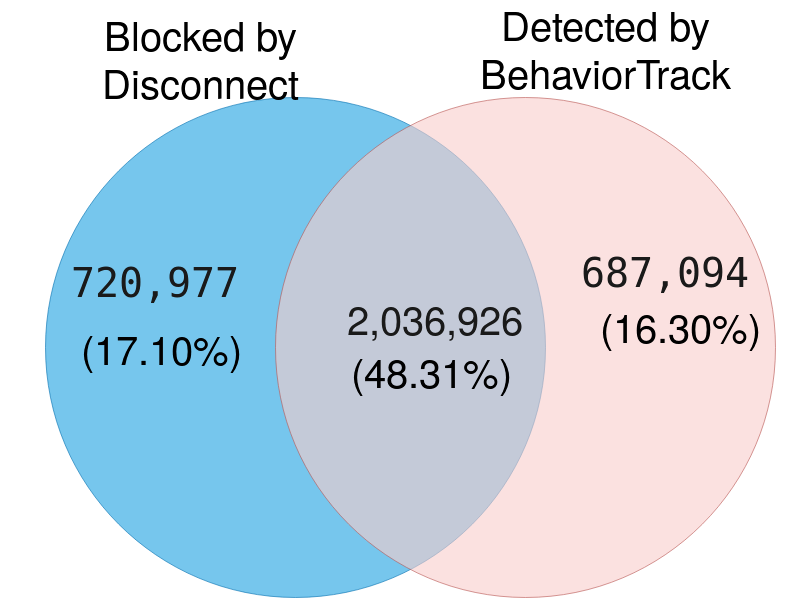}}
\caption{ {\normalfont}
Effectiveness of filter lists at detecting trackers on $4,216,454$ third party requests from \totalpagessucc\ pages.} 
\label{fig:comparing} 
\SHORTEN
\end{figure}

{\bf Measuring tracking requests}
We apply filter lists on requests to detect which requests are 
blocked by the lists,
as it has been done in previous works~\cite{Engl-Nara-16-CCS}. 
We then use the filter lists to classify follow-up third-party requests that would have been blocked 
by the lists. This technique has been extensively used in the previous works 
\cite{Engl-etal-18-PETS,Iord-etal-18-IMC,Laui-etal-17-NDSS,Ikra-etal-17-PETS} (for more details, see Table~\ref{tab:ELEPusage} in the Appendix).
We classify a request as blocked if it matches one of the conditions: 

\begin{itemize}
\item 
the request directly matches the list. 
\item 
the request is a consequence of a redirection chain where an earlier request was blocked by the list.
\item 
the request  is loaded in a third-party content (an iframe) that was blocked by the list (we detect this case by analyzing the referrer header).
\end{itemize}

Figure~\ref{fig:comparing} provides an overview of 
third party requests 
blocked by filter lists or detected as tracking requests according to \ourmethod.
Out of all $4,216,454$ third party requests  in the \emph{full dataset}, 
$2,558,921$ (60.7\%) requests were blocked by \ELEP,  
$2,757,903$ (65.4\%) were blocked by Disconnect, and 
$2,724,020$ (64.6\%) were detected as performing tracking by {\ourmethod}.

\begin{table*}[!th]
    \centering
    \begin{tabular}{p{1.3cm}|p{1.2cm}|p{1.4cm}|p{1.3cm}|p{1.4cm}|p{1.3cm}|p{1.4cm}}
    \textbf{Filter list(s)}  & \# missed requests  & \% of 4.2M  third-party requests & \% of 2.7M tracking requests & \# domains responsible for missed requests & \# trackers follow up & \#  effective missed tracking requests \\ \hline
    \ELEP 	&  $826,622$  & 19.60\% & 25.22\% & $5,136$ &  $118,314$ & $708,308$ \\ 
    \disc 	& $687,094$ & 16.30\% & 30.34\% & $6,189$ & $46,285$  &  $640,809$ \\ 
    \end{tabular}
    \caption{Overview of third-party requests missed by the filter lists and detected as tracking by {\ourmethod}.}
    \label{tab:overrequests}
    \vspace{-0.7cm}
\end{table*}{}

\textbf{Requests blocked only by filter lists: }
Figure~\ref{fig:comparing}  shows 
 that 
 {\ELEP}  block $661,523$ ($15.69$\% out of $4,216,454$) requests that were 
 not detected as performing tracking by {\ourmethod}. 
 These requests originate from 
 $2,121$ 
 unique third party domains. 
 \disc\ blocks $720,977$ ($17.10\%$) 
requests not detected by \ourmethod. These requests 
originate from $1,754$ distinct third party domains.
 
 These requests are missed by \ourmethod\ because they do not contain 
 any {\id}. Such requests may contain other non user-specific cookies (identical across 
 two machines, see Sec. \ref{DetectingcookieID}), however we assume that such cookies are not used for tracking. 
 \ELEP\ and {\disc} block these requests most likely because
 they are known for providing analytics or advertising services, 
or because they  perform other types  of tracking through scripts 
 such as fingerprinting, which is out of the scope of our study.

\subsection{Tracking  missed by the filter lists}

Table~\ref{tab:overrequests} gives an overview of third-party requests missed by \ELEP\ 
and \disc\ filter lists and 
 detected by {\ourmethod} as performing tracking. 
The number of third party domains involved in 
tracking detected only by $\ourmethod$ (e.g., $6,189$ for \disc) is significantly higher than 
those only detected  
by filter lists (e.g., $1,754$ for \disc\ as reported earlier in this section).
We define the term \textit{ trackers follow up} as the requests using  identifying cookies set by previous requests blocked by the filter lists (note that our crawler is stateful).
As a result, by simulating the blocking behavior of the filter lists, 
these cookies should be blocked 
and not  included in the analysis of the following requests.
Consequently, the follow up requests should not be categorized as tracking requests.\\
By further analyzing the requests only detected as tracking by {\ourmethod} and missed by {\ELEP}, 
we found that $118,314$  requests ($14.31$\% of the requests detected only by {\ourmethod})  are 
trackers follow up.
Similarly,
we found that $46,285$  requests  ($6.73$\% of the requests detected only by {\ourmethod})  missed by {\disc} are trackers follow up.
We exclude these requests from the following analysis
and we further analyze the remaining $708,308$ requests missed by {\ELEP} and the $640,809$ missed by {\disc}.

\ourmethod\ detects 
all kind of trackers including the less popular ones that are under the bar of detection of filter list.
Because less popular trackers are less prevalent, they generate fewer requests 
and therefore remain unnoticed by filter lists. 
This is the reason why we detect a large number  of domains responsible for tracking. 
\subsubsection{Tracking enabled by useful content}
\begin{table}[!h]
    \centering
    \begin{tabular}{l|c|c}
    \textbf{Content type} &  \textbf{Missed by \ELEP} & \textbf{Missed by Disconnect} \\ \hline
    script  &   33.38\% &  35.27 \% \\ 
    big images  &  20.62\% &   21.73 \%  \\ 
    text/html  &   13.77\% &   14.73 \%  \\ 
    font  &   8.79\% &  0.09 \%  \\ 
    invisible images  &  6.68\% & 12.21 \% \\ 
    stylesheet  &  6.17\% &   3.05 \%  \\ 
    application/json  &   4.00\% &  4.83 \% \\ 
    others & 6.59\% & 8.12\% \\ 
    \end{tabular}
    \caption{Top content type detected by {\ourmethod} and not by filter lists on the $708,308$ requests missed by {\ELEP} and the $640,809$ missed by {\disc}}
    \label{tab:contentelep}
    \vspace{-0.7cm}

\end{table}{}

We analyzed the type of content provided by the remaining tracking requests.
Table \ref{tab:contentelep} presents the top content types 
used for tracking and not blocked by the filter lists.
We refer to images with dimensions larger than 50×50 pixels as Big images.
These kinds of images, texts, fonts and even stylesheets 
are used for tracking.
The use of these types of contents is essential for the proper functioning of the website.
That makes the blocking of responsible requests by the filter lists impossible.
In fact, the lists are explicitly allowing content from some of these trackers to avoid the breakage of the website,
as it's the case for \domain{cse.google.com}.

\begin{table}[]
    \centering
    \begin{tabular}{l|c|c}
    \textbf{Service category} &  \ELEP & \disc \\ \hline
    Content Servers  &   23.33 \% &  23.33 \%\\ 
    Social Networking  &   16.67 \% & 0.00\%  \\ 
    Web   Ads/Analytics  &    13.33 \% &  23.33 \% \\ 
    Search Engines/Portals  &    13.33 \% &  23.33 \% \\ 
    Technology/Internet  &   13.33 \% &  10.00 \% \\ 
    Consent frameworks  &   3.33 \% & 3.33 \%  \\
    Travel  &   3.33 \% &   3.33 \% \\ 
    Non Viewable/Infrastructure  &    3.33 \% &  0.00\%   \\ 
    Shopping  &   3.33 \% &   3.33 \% \\ 
    Business/Economy  &    3.33 \% &  6.67 \% \\ 
    Audio/Video Clips  &    3.33 \% &  0.00\%  \\ 
    Suspicious  & 0.00\%  &   3.33 \% \\ 

    \end{tabular}
    \caption{Categories of the top 30 tracking services detected by {\ourmethod} and missed by the filter lists.}
    \label{tab:servcat}
    \vspace{-0.7cm}
\end{table}{}

We categorized the top $30$ third party services
not blocked by the filter lists but detected by {\ourmethod} as performing tracking using Symantec's WebPulse Site Review \cite{categorization}. Unlike in previous sections, where we analyzed the  
$2^{nd}$-level TLD, such as \domain{google.com}, here we report on full domain names, 
such as \domain{cse.google.com}. That gives more information about the service provided.
New domains such as \domain{consensu.org} are not categorized properly so we manually added a new category called “Consent frameworks” 
to our categorization for such services.
Table \ref{tab:servcat} represents the results of this categorization.
Web   Ads/Analytics represents 13.33\% of the services missed by {\ELEP} and 23.33\% of those missed by {\disc}.
However, the remaining services are mainly categorized as 
content servers, search engines and other functional categories. 
They are tracking the user, 
but not blocked by the lists. This is most likely not to break the websites.

\subsubsection{Why useful content is tracking the user}
\textbf{Tracking enabled by a first party cookie: }
A cookie set in the first party context can be considered as a third party cookie
in a different context.
For example, a  \domain{site.com} cookie is a first party cookie 
when the user is visiting \domain{site.com}, 
but 
it becomes a third party when the user is visiting a different website that includes 
content from \domain{site.com}. 
Whenever a request is sent to a domain, say \domain{site.com}, the browser
automatically attaches all the cookies that are labeled with \domain{site.com} to 
this request.

For example, when a user visits \domain{google.com}, a first party \id\ is set. Later on, when 
a user visits \domain{w3school.com},
a request 
is sent to the service \domain{cse.google.com} (Custom Search Engine by Google).
Along with the request, Google's {\id} 
is sent to  \domain{cse.google.com}.
The filter list 
cannot block such a request, and 
is incapable of removing the first party tracking cookies from it. 
In our example, filter lists do not block 
the requests sent to  \domain{cse.google.com}
on 329 different websites.
In fact, blocking \domain{cse.google.com} breaks the functionality of the website.
Consequently, 
an \id\ 
is sent to the \domain{cse.google.com},
allowing it to track the user across websites.

By analyzing the requests missed by the lists,
we found that this behavior explains a significant amount of missed requests: 
44.61\% requests ($316,008$  out of  $708,308$)  missed by {\ELEP}
and 
32.00\%  requests ($205,088$ out of $640,809$) missed by {\disc}
contain cookies initially set in a first party context.

\textbf{Tracking enabled by large scope cookies.}
A cookie set with a $2^{nd}$-level TLD domain 
can be accessed by all its subdmains.
For example, 
a third party \domain{sub.tracker.com} sets a cookie in the user browser
with \domain{tracker.com} as its domain.
The browser 
sends this cookie to another subdomain of \domain{tracker.com} 
whenever a request to that subdomain is made. 
As a  result of this practice,
the {\id} set by a tracking subdomain with $2^{nd}$-level TLD domain
is sent to all other subdomains, even the ones serving useful content.

Large scope cookies are extremely prevalent among requests missed by the 
filter lists. 
By analyzing the requests missed by the lists,
we found that $77.08$\% out of 22,606 
third-party cookies used in the requests missed by {\ELEP} and
$75.41$\% out of 24,934 
cookies 
used in 
requests missed  by \disc\ 
were set with a $2^{nd}$-level TLD domain (such as \domain{tracker.com}).

\subsection{Panorama of missed trackers}

To study the 
effectiveness of 
\ELEP\ and  \DISC\ combined,
we compare requests blocked by these filter lists with requests detected by \ourmethod\
as tracking according to the classification from Figure~\ref{fig:categorizationoverview}. 
These results are based
on the dataset of $4,216,454$ third-party requests collected from {\totalpagessucc} pages of $8,744$ domains.

Overall, 
$379,245$  requests originating from 
9,342 services (full third-party domains)
detected by \ourmethod\ 
are not blocked by \ELEP\ and {\disc}. 
Yet, these requests are performing at least one type of tracking, 
they represent 9.00\% of all $4,2$M third-party requests and appear in 68.70\% of websites.

We have detected that the  $379,245$  requests detected by \ourmethod\ perform 
at least one of the tracking behaviors presented in Figure \ref{fig:categorizationoverview}.
Table~\ref{tab:categoriesmissed} 
  shows the distribution  of tracking behaviors detected by \ourmethod.
\begin{table}[]
    \centering
    \begin{tabular}{l|c}
    \textbf{Tracking behavior} & \textbf{Prevalence} \\ \hline
     Basic tracking  &  83.90\%  \\ \hline
    Basic tracking initiated by a tracker  &  13.50\% \\ \hline
    First to third party cookie syncing  &  1.42\% \\ \hline
    Analytics  &  1.00\% \\ \hline
    Third to third party cookie syncing  &  0.09\%  \\ \hline
    Third party cookie forwarding  &  0.08\%  \\ \hline
    \end{tabular}
    \caption{Distribution of tracking behaviors in the $379,245$ requests missed by \ELEP\ and Disconnect.}
    \label{tab:categoriesmissed}
    \vspace{-0.7cm}

\end{table}{}
We notice that the most privacy-violating behavior that 
includes setting, sending or syncing third-party cookies is represented by the 
basic tracking that is  present in  
(83.90\%)  of missed requests.

Table~\ref{tab:finalresults} in Appendix  presents the top 15 
domains detected as trackers and missed by the filter lists.
For each domain, we extract its  category, owners and country of registration 
using the whois library~\cite{whois} and manual checks.
We also manually analyzed all the cookies associated to tracking:
out of the $15$ presented domains,
$7$ 
are  tracking the user using persistent first party cookies. 
The cookies of the search engine Baidu 
expires within 68 years,
 whereas the cookies associated to Qualtrics, an experience management company, 
expires  in 100 years.

We found that content from  \domain{code.jquery.com},  \domain{s3.amazonaws.com}, and \domain{cse.google.com} 
are explicitly allowed by the filter lists  on a list of predefined first-party websites 
to avoid the breakage of these websites.
We identified
\domain{static.quantcast.mgr.consensu.org} by {IAB Europe} 
that rightfully should not be blocked
because they provide useful functionality for GDPR compliance. 
We detect that the 
cookie values seemed to be unique identifiers, but are set without expiration date, which means
such cookies will get deleted when the user closes her browser. Nevertheless, it is known that users 
rarely close browsers, and more importantly, it is unclear why a consent framework system sets
identifier cookies even before the user clicks on the consent button (remember that 
we did not emit any user behavior, like clicking on buttons or links during our crawls).  

We identified tag managers -- 
these tools are designed to help Web developers to manage marketing and tracking tags
 on their websites and can't be blocked not to break the functionality of the website. 
 We detected that two such managers,  \domain{tags.tiqcdn.com}  by Tealium 
 and \domain{assets.adobedtm.com} by Adobe track users cross-sites, but 
 have an explicit exception in EasyList.

\section{Are browser extensions effective at blocking trackers?}
In this Section,
we analyze how effective are the popular privacy protection extensions in blocking the privacy leaks detected by {\ourmethod}.
We study the following extensions: Adblock~\cite{AdBlock}, Ghostery~\cite{ghostery}, Disconnect~\cite{Disconnect}, 
and Privacy Badger~\cite{privacyBadger}.
The latest version of 
uBlock Origin 1.22.2 is not working correctly with 
OpenWPM under Firefox 52, which is the latest version of Firefox running on OpenWPM that supports both web extensions and stateful crawling.
Hence we didn't include uBlock Origin in our study.

We performed simultaneous stateful Web measurements of the Alexa top 10K websites using OpenWPM in November 2019 from servers located in  France.
For each website, we visit the homepage and 2 randomly chosen links on the homepage from the same domain. Selection of links was made in advance.\\
We consider the following measurement scenarios: 
\begin{enumerate}
\item Firefox with no extension. 
\item Firefox with Adblock 3.33.0 (default settings).
\item Firefox with Ghostery 8.3.4 (activated blocking).
\item Firefox with Disconnect 5.19.3 (default settings).
\item Firefox with Privacy Badger 2019.7.11 (trained 
on the homepage  and 2 random links from this homepage for the  
top 1,000 Alexa websites). 
\item Firefox with all previous extensions combined.
\end{enumerate}

Out of $30,000$ crawled pages, $25,485$ were successfully loaded by all the crawls. The analysis in the following is done on this set of pages.

\begin{figure}[t]
\centering
\includegraphics[width=0.35\textwidth]{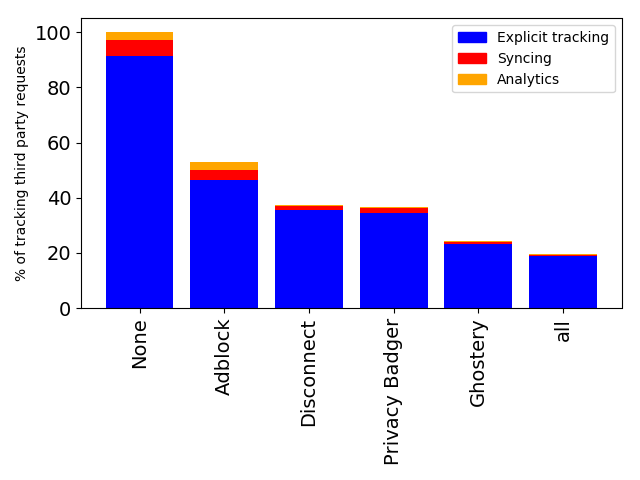}
\caption{\label{fig:exten} 
Percentage of third party requests allowed by privacy protecting browser 
extensions 
out of 2,924,480 tracking requests. 
}
\SHORTEN
\end{figure}

Figure \ref{fig:exten} represents the effectiveness of the extensions in blocking the tracking requests detected by {\ourmethod}.
Our results show that Ghostery is the most efficient among them.
However, it still fails to block $24.38\%$ of the tracking requests.
All extensions miss trackers in the three classes,
However, Disconnect and Privacy Badger have an efficient Analytics blocking  mechanism: 
they are missing Analytics behavior on only 0.36\% and 0.27\% of the  pages respectively.
Most tracking requests missed by the extensions are performing Explicit tracking.\\
\textbf{Conclusion:}
Similarly to Merzdovnik {\al} \cite{Merz-etal-17-EuroSnP}, we show that tracker blockers 
(Disconnect, Ghostery and Badger) are more efficient than adblockers (Adblock) in blocking tracking behaviors.
However, all studied extensions miss at least $24.38\%$ of the tracking detected by {\ourmethod}.
This shows that even though the extensions reduce the amount of tracking performed,
they do not solve the 
problem of protecting users from tracking.

\section{Discussion}
\label{sec:discussion}

Our results show that there are numerous problems in the cookie-based third party 
tracking. In this Section, we discuss these problems 
with respect to different actors.

{\bf Browser vendors}. 
We observed that 
first party cookies can be exploited in a third party context 
to perform cross site tracking.
In its Intelligent Tracking Prevention 2.0 introduced in 2018,
Safari allowed cookies to be used in a third party context only in the first  24 hours of the cookie lifetime.
Such time frame could be limited even further, however this 
approach requires rigorous testing with end users. 
Other browser vendors 
should follow
Safari and prohibit the usage of  cookies 
in a third party context.

{\bf Web standardization organizations.} 
While third-party content provides 
useful features to the website, 
it is also capable of tracking users. 
We have shown that third party domains serving functional content such as 
Content Servers or Search Engines may 
track the user with identifier cookies. 
We have noticed that we detect such tracking because the domain 
behind such functional content does not set but only receives identifier cookies 
that are already present in the browser and were initially set 
with the $2^{nd}$-level domain as host,
which makes the cookie accessible by all subdomains.
Even if the tracking is not intentional, 
and the domain is not using the {\id} it receives to create user's profile currently,
this cookie leakage is still a privacy concern that could be exploited by the service anytime.
We therefore believe that Web standardization bodies, such as the W3C, could 
propose to limit the scope of the cookies and not send it to all the subdomains.

 {\bf Supervisory bodies.}
 When a supervisory body, such as a Data Protection Authority in the EU, has to investigate 
 and find the responsible party for the tracking happening on a website, 
 it is a very complex task to 
 identify who is liable for setting or sending the 
 {\id}. 
 In our work, we have identified \emph{tracking initiators} -- 
 third party domains that only redirect or include other domains that perform tracking. 
 Such tracking initiators, that we detected on 11.24\% of websites, are partially liable 
 for tracking. Another example are CDNs, we have observed that requests or responses for 
 fetching a jQuery library 
 from  \texttt{code.jquery.com} contains {\id}s. 
We found that
it is the  Cloudflare CDN that inserts a cookie named \texttt{\_\_cfduid} 
 into its traffic in order ``to identify malicious visitors to their  Customers’ websites''.

{\bf Conclusion.} 
Our work raises 
numerous concerns in the area of tracking detection  
and privacy protection of Web users. We believe that our work can be used to 
improve existing tracking detection approaches, but nevertheless various actors  need to revise their practices when 
it comes to the scope and usage of cookies, and third parties 
should exclude third party tracking from the delivery of functional website content.

%% file: relatedwork.tex
\section{Related Work}
\label{sec:relatedwork}

In this Section, 
we  first provide an  overview of 
previous works on measuring invisible pixels. We then examine 
state-of-the-art techniques to detect online tracking: 
behavior-based techniques and methods leveraging the filter lists.

\subsection{Invisible pixels, known as web bugs}
Invisible pixels are 
extensively studied starting from 2001~\cite{Nich-01,Alsa-mart-02-PETS, Mart-etal-03-ACM, Dobi-10-IFIP,Ruoh-Lepp-18}.
Invisible pixels, called ``web bugs'' in previous works, were primarily used to set and send third-party cookies 
attached to the request or response when the browser fetches such image.
In 2003, Martin et al.~\cite{Mart-etal-03-ACM} 
found that 58\% of the 84 most popular websites and 36\% of 289 random websites contain at least one web bug.
In 2002, Alsaid and Martin \cite{Alsa-mart-02-PETS} deployed a tool (\emph{Bugnosis})
to detect the web bugs.
The main goal of the tool was to raise awareness among the public.
It was used by more than 100,000 users.
However, it was only generating warning messages without actively blocking the bugs and  was only supported by Internet Explorer 5, that is deprecated today.
Dobias \cite{Dobi-10-IFIP} showed 
 that 
 web bugs lead to new privacy threats, such as 
fingerprinting.

Ruohonen and Leppänen~\cite{Ruoh-Lepp-18} studied the presence of 
invisible pixels in  Alexa’s top 500 
websites. 
They showed that invisible pixels are still widely used.
Differently from our work, where we detect all effectively delivered images from the response headers, 
the authors analyze the source code of landing HTML page and 
extract images from the <img>  tag.
Such a method misses 
an important number of images that are  dynamically loaded.

The significant number of studies on invisible pixels 
shows that it is a well known problem.
The goal of our study is different:
\emph{we aim to use invisible pixels that are still  widely present on the Web 
to detect different tracking behaviors and  collaborations}.

\subsection{Detection of online tracking}
{\bf Detection of trackers by analysing behavior: }
Roesner et al.~\cite{Roes-etal-12-NSDI} and Lerner et al.~\cite{Lern-etal-16-USENIX} were 
the first to analyze trackers based on their behavior. 
They have proposed a classification of tracking behaviors 
that makes a distinction between analytics and cross-domain tracking. 
We, however, 
propose a more \emph{fine-grained classification of tracking behaviors} 
that includes not only previously known behaviors, but also 
specific categories of cookie sharing and syncing 
(see Section~\ref{sec:results}).
Yu {\al} \cite{Yu-etal-16-WWW} identify trackers by detecting unsafe data 
without taking into account the behavior of the third party domain and the 
communications between trackers.

Previous studies ~\cite{Acar-etal-14-CCS,Olej-etal-14-NDSS,Engl-Nara-16-CCS,Bash-etal-16-USENIX,papa-etal-19-www} measured cookie syncing on websites and users.

Olejnik \etal~\cite{Olej-etal-14-NDSS} considered 
cookies with sufficiently long values to be identifiers.
If 
such identifier is shared between domains, 
then it is classified as 
cookie syncing. Additionally, 
Olejnik \etal~\cite{Olej-etal-14-NDSS}
studied the 
case of \domain{doubleclick.net} to detect cookie syncing based on the URL patterns. 
In our study, we show that domains are using more complex techniques to store and share identifier cookies.
%
We base our technique for detecting identifier cookies on the work 
of Acar \etal~\cite{Acar-etal-14-CCS}, and Englehardt and 
Narayanan~\cite{Engl-Nara-16-CCS}, 
who only checked for the identifiers that are stored and shared 
in a clear text. 
In our work, we 
detect more cases of cookie synchronization
because we detect 
encoded cookies and even encrypted ones in 
the case of \domain{doubleclick.net}.
Bashir \etal~\cite{Bash-etal-16-USENIX}, used retargeted ads to detect cookie syncing. 
To detect these ads, authors filtered out all images with dimensions lower than 50×50 pixels,
then they studied the information flow leading to these images.
Which limit their study to chains resulting to a retargeted ad.
In our work, we analyse all kind of requests.

Papadopoulos \etal~\cite{papa-etal-19-www} used a year-long dataset
from real mobile users 
to study cookie syncing. 
The authors 
detect not only syncing done through clear text,
but encrypted cookie syncing as well.
Hence, they cover DS, PC and ES sharing techniques detected by {\ourmethod} (see Figure \ref{fig:cookiesharing}),
but they miss the 
remaining techniques that represent $39.03$\% of the 
cookie sharing 
that we detect.
Moreover, 
they only focus on cookie syncing, 
while we conduct a more in-depth study of different tracking 
behaviors extracted from invisible pixels dataset,
and we compare our tracking detection to filter lists 
and the most 
popular privacy extensions.

%% file: consumerlists.tex
{\bf Detection of trackers with filter lists:}
To 
detect domains related to tracking or 
advertisement, most of the previous studies \cite{Engl-Nara-16-CCS,Bash-etal-16-USENIX,Laui-etal-17-NDSS,Raza-etal-18-NDSS,Ikra-etal-17-PETS,Engl-etal-18-PETS,Bash-Wils-18-PETS,Bash-etal-18-IMC,Iord-etal-18-IMC} rely on 
filter lists, 
such as 
EasyList~\cite{easylist} and EasyPrivasy~\cite{easyprivacy} (EL\&EP) 
that became the \emph{de facto} approach to detect 
trackers. 
From the 
last three years alone, 
we identified \emph{9 papers that rely on \ELEP} to detect  third-party tracking and advertising
(see Table~\ref{tab:ELEPusage} in the Appendix). 

Englehardt and Narayanan~\cite{Engl-Nara-16-CCS} seminal work on measuring trackers on 1 million 
websites 
relies on \ELEP\ as a ground truth to detect requests sent to trackers and ad-related domains.
Three papers by Bashir et al.~\cite{Bash-etal-16-USENIX,Bash-etal-18-IMC,Bash-Wils-18-PETS} customize \ELEP\ to detect $2^{\mathit{nd}}$-level domains of tracking and ad companies: to eliminate false positives, a domain is considered 
if it appears more than 10\% of the time in the dataset. 
Lauinger et al. \cite{Laui-etal-17-NDSS} use \ELEP\ to identify advertising and tracking content in order to detect what content has included outdated and vulnerable JavaScript libraries in Web applications. 
Razaghpanah et al.~\cite{Raza-etal-18-NDSS} use EasyList as an input to their classifier to identify advertising and tracking domains in Web and mobile ecosystems.
Ikram et al.~\cite{Ikra-etal-17-PETS}  analysed how many tracking JavaScript libraries are blocked by \ELEP\ 
on 95 websites. 
Englehardt et al.~\cite{Engl-etal-18-PETS} apply \ELEP\ 
on third-party leaks caused by invisible images in emails. 
Iordanou \etal~\cite{Iord-etal-18-IMC} rely on \ELEP\ as a ground truth for detecting ad- and tracking-related third party requests.
Only one work by Papadopoulos et al.~\cite{Papa-etal-17-IMC} uses Disconnect list~\cite{DisconnectList} to detect tracking domains. 
To the best of our knowledge, \emph{we are the first to compare  
 the behavior-based detection method to 
 filter lists 
 extensively used in the literature.}

\textbf{Effectiveness of filter lists: }
Merzdovnik et al \cite{Merz-etal-17-EuroSnP} studied the effectiveness 
of the most popular tracking blocking extensions.
They evaluate how many third party requests are blocked by each extension.
In their evaluation, they don't distinguish tracking third party requests from non tracking ones, which affects their evaluation.
In our work, we detect trackers using a behavior-based detection method
and then we evaluate
how much of these trackers are blocked.
Das et.al \cite{Das-etal-18-CCS} studied the effectiveness of filter lists against tracking scripts
that misuse sensors on mobile.
They show that filter lists fail to block the scripts  that access the sensors.
We instead evaluate effectiveness of filter lists against third party requests in web applications that contain {\id}s.

%% file: limitation.tex
%% file: conclusion.tex
\section{Conclusion}
Web tracking remains an important problem for the privacy of Web users. 
Even after the General Data Protection Regulation (GDPR) came in force 
in May 2018, third party companies continue tracking users with various sophisticated 
techniques based on cookies without their consent. According to our study, 
91.92\% of websites incorporate at least one type of cookie-based tracking.  

In this paper, we define a new classification of Web tracking behaviors,
thanks to a large scale study of invisible pixels collected from {\totalpagessucc} webpages. 
We then applied our classification to the full dataset 
which allowed us to uncover different relationships between domains.
The redirection process and the different behaviors that a domain can adopt 
are an evidence of the complexity of these relationships.
We show that even the most popular consumer protection lists 
and browser extensions 
fail to detect these complex behaviors.
Therefore, 
behavior-based tracking detection should be more widely adopted.

\section{Acknowledgment}
First of all, we would like to thank the valuable comments and
suggestions of the anonymous reviewers of our paper. 
This work is supported by ANSWER project PIA FSN2 (P159564-266178\textbackslash DOS0060094).

%% file: appendix.tex
\section{Appendix}
\label{sec:appendix}
\subsection{Detecting identifier sharing} \label{sharingapp}

\textbf{GA sharing: } \domain{Google-analytics} serves invisible pixels on 69.89\% of crawled domains 
as we show in Figure~\ref{fig:resp_dom}.
By analyzing our data, we detect that 
the cookie set by \domain{google-analytics} script is of the form GAX.Y.Z.C,  
while the \emph{identifier cookies} sent in the parameter value to \domain{google-analytics} is actually Z.C.
This case is not detected by the previous 
cookie syncing detection techniques for two reasons. 
First, "." is not considered as a delimiter. Second, even if it was considered as a delimiter, it would create 
a set of values \{GAX, Y, Z, C\} which are still different than the real value Z.C used as an identifier by \domain{google-analytics}. 

\textbf{Base64 sharing: } 
When a domain wants to share its 
\emph{identifier cookie} with \domain{doubleclick.net},
it should encode it in base64 before sending it 
\cite{Doubleclick}.
For example, when \domain{adnxs.com} sends a request to \domain{doubleclick.net}, 
it includes a random string into a URL parameter. This string is 
the base64 encoding of the value of the cookie set by \domain{adnxs.com} in the user's browser.\\
\sloppy{}
\textbf{Encrypted sharing: } 
When \domain{doubleclick.net} wants to share its \emph{identifier cookie} with some other domain, 
it encrypts the cookie before sending it, 
which makes the detection of the identifier cookie sharing impossible. 
Instead, we rely on the semantic defined by doubleclick to share this identifier ~\cite{Doubleclick}. 

Assume that \domain{doubleclick.net} is willing to share an identifier cookie with \domain{adnxs.com}.
To do so, Doubleclick requires that the content of  \domain{adnxs.com} includes an image tag, 
pointing to a 
URL that contains  \domain{doubleclick.net} as destination
and a parameter \textit{google\_nid}. 
Using the value of the  \textit{google\_nid} parameter, \domain{doubleclick.net} get to know
that \domain{adnxs.com}  was the initiator of this request. 
Upon receiving such request, \domain{doubleclick.net} sends a redirection response 
pointing to a URL that contains \domain{adnxs.com} as destination 
with encrypted  \domain{doubleclick.net}'s cookies in the parameters. 
When the browser receives this response, it redirects to \domain{adnxs.com}, 
who now receives encrypted   \domain{doubleclick.net}'s cookie. 

We detect such behavior by detecting requests to \domain{doubleclick.net} with 
 \textit{google\_nid} parameter and analysing the following redirection. 
 If we notice that the redirection is set to a concrete domain, for example \domain{adnxs.com}, 
 we conclude that \domain{doubleclick.net} has shared its cookie with this domain. 

\subsection{Additional results}

\begin{figure}[!htb]
\centering
\includegraphics[width=0.35\textwidth]{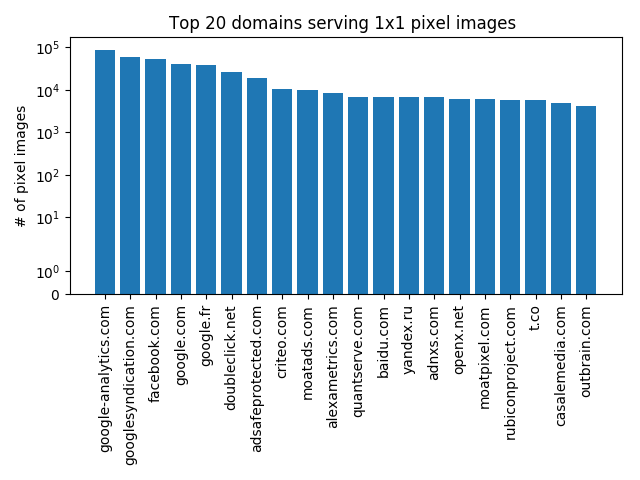}
\caption[lenght]{Top 20 domains responsible for serving invisible pixels}
\label{fig:resp_dom}
\end{figure}

Figure \ref{fig:resp_dom} represents the Top 20 domains involved in invisible pixels inclusion in the $\totaldomainssucc$  domains.

Table \ref{tab:host} presents the top 10 domains using their cookie key to store the identifier.
\begin{table}[!htbp]
    \centering
    \begin{tabular}{l|c}
    \textbf{Host} & \textbf{\# cookies instances} \\ \hline
    lpsnmedia.net & 583\\ \hline
    i-mobile.co.jp & 223\\ \hline
    rubiconproject.com & 83\\ \hline
    justpremium.com & 72\\ \hline
    juicyads.com & 64\\ \hline
    kinoafisha.info & 64\\ \hline
    aktualne.cz & 63\\ \hline
    maximonline.ru & 61\\ \hline
    sexad.net & 47\\ \hline
    russian7.ru & 45\\ \hline
    \end{tabular}
    \caption{Top 10  domains storing the identifier as key.}
    \label{tab:host}
    	\SHORTEN
\end{table}

\begin{figure}[!htb]
\centering
\includegraphics[width=0.4\textwidth]{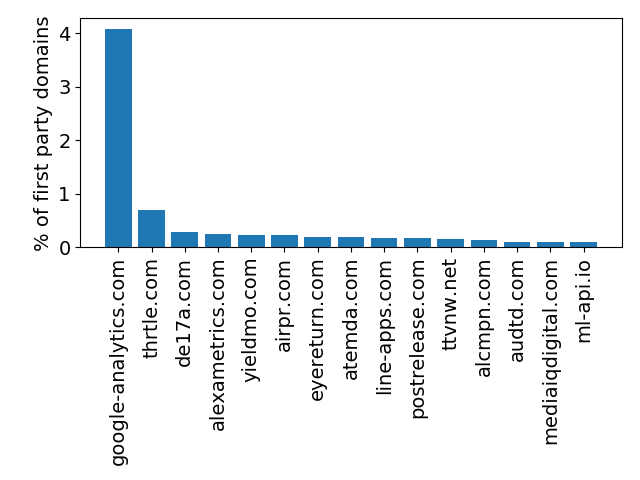}
\caption{\label{fig:receivers_transfer_cookie_no_fp}Third party cookie forwarding : Top 15 receivers in $\totaldomainssucc$  domains. }
\end{figure}

Figure \ref{fig:receivers_transfer_cookie_no_fp} represents the Top 15 third parties receiving the identifier cookies. Google-analytics is the top domain receiving identifiers in over 4\% of the visited websites.
Table \ref{tab:gafrwd} presents the top 10 third parties sharing their identifiers with google-analytics.

\begin{table}[htbp!]
\centering
\begin{tabular}{l|c}
\hline
Third parties & \# requests \\ \hline
adtrue.com  &  298  \\ \hline
google.com  &  123 \\ \hline
architonic.com   &  120 \\ \hline
bidgear.com  &  80\\ \hline
akc.tv  &  76  \\ \hline
insticator.com   &  73 \\ \hline
coinad.com &  64  \\ \hline
performgroup.com &  52 \\ \hline
chaturbate.com  &  47  \\ \hline
2mdnsys.com &  40 \\ \hline

\end{tabular}
\caption{\label{tab:gafrwd}Third party cookie forwarding; Top 10 third parties forwarding cookies to \domain{google-analytics}. }
\vspace{-0.5cm}
\end{table}

\begin{figure}[!htbp]
\centering
\includegraphics[width=0.4\textwidth]{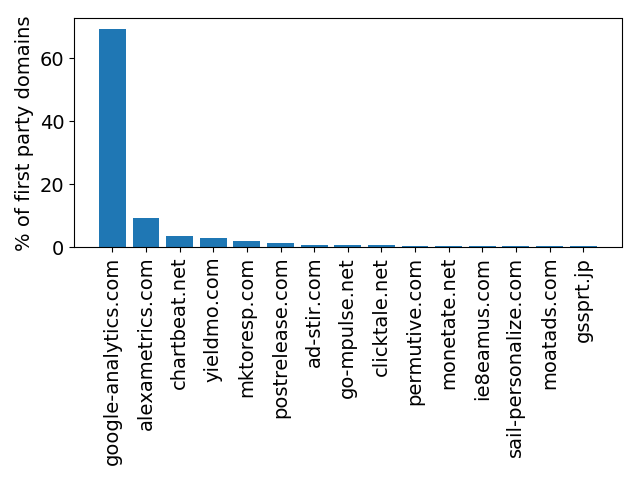}
\caption{\label{fig:recanaly} Analytics: Top 15 receivers  in the $\totaldomainssucc$  domains. }
\SHORTEN
\end{figure}

Figure \ref{fig:recanaly} presents the top 15 analytics domains in our dataset of 8,744 domains.

\begin{table*}[h!]
\centering
\begin{tabular}{p{3.3cm}|p{1.7cm}|p{2.1cm}|p{1.4cm}|p{2.3cm}|p{4.1cm}|p{0.9cm}}
\hline
\multicolumn{6}{l}{\hspace{4cm} \bf  Tracking enabled by a first party cookie}  \\
\hline
Full domain			 	& Prevalence of tracking in first-parties & Cookie name  	& Cookie \hspace{0.5cm} expiration &Category & 	Company & Country  \\ \hline

code.jquery.com  &   756  ( 8.65 \%)  &  \_\_cfduid & 1 years & Technology/Internet &jQuery Foundation & US \\ \hline

s3.amazonaws.com  &  412  ( 4.71 \%)  & s\_fid & 5 years & Content Servers &  Amazon & US \\ \hline
\hline

ampcid.google.com  &  282  ( 3.23 \%) & NID & 6 months &Search Engines & Google LLC  &  US \\ \hline
cse.google.com  &  307  ( 3.51 \%) &  NID & 1 year &Search Engines & Google LLC  &  US \\ \hline

use.fontawesome.com  &  221  ( 2.53 \%) & \_\_stripe\_mid &  1 years &Technology/Internet & WhoisGuard Protected & \_  \\ \hline

siteintercept.qualtrics.com  &  99  ( 1.13 \%) & t\_uid & 100 years &Business/Economy& Qualtrics, LLC & US \\ \hline

push.zhanzhang.baidu.com  &  98  ( 1.12 \%) & BAIDUID  &  68 years &Search Engines& Beijing Baidu Netcom  Science Technology Co., Ltd.  &  CN \\  \hline

\multicolumn{6}{l}{\hspace{4cm} \bf Tracking enabled in a third party context}  \\
\hline
assets.adobedtm.com  &  427  ( 4.88 \%) & \_gd\_visitor  & 20 years  &Technology/Internet& Adobe Inc. & US \\ \hline
\
yastatic.net  &  303  ( 3.47 \%)  & cto\_lwid & 1 year &Technology/Internet& Yandex N.V.  &  RU \\ \hline
s.sspqns.com  &  278  ( 3.18 \%)  & tuuid & 6 months &Web   Ads/Analytics& HI-MEDIA & FR \\ \hline
tags.tiqcdn.com  &  276  ( 3.16 \%) & utag\_main& 1 year & Content Servers& Tealium Inc  & US \\ \hline
cdnjs.cloudflare.com  &  206  ( 2.36 \%)  & \_\_cfduid & 1 year & Content Servers& Cloudflare & US \\ \hline

\hline 
static.quantcast.mgr.\hspace{0.5cm}consensu.org  &  157  ( 1.80 \%)  & \_cmpQc3pChkKey &  Session & Consent \hspace{0.4cm} frameworks & IAB Europe  &  BE \\ \hline

a.twiago.com  &  133  ( 1.52 \%) & deuxesse\_uxid &  1 month &Office/Business \hspace{0.4cm} Applications&  REDACTED FOR PRIVACY & \_ \\ \hline
g.alicdn.com  &  121  ( 1.38 \%) & \_uab\_collina & 10 years & Content Servers & Alibaba Cloud Computing Ltd.  &  CN \\ \hline

\end{tabular}
\caption{\label{tab:finalresults} Top 15  domains missed by \ELEP\ and \DISC\ but detected 
by \ourmethod\ to perform tracking. }
\SHORTEN
\end{table*}

Table~\ref{tab:finalresults}  presents the  top 15 
domains detected as trackers and missed by the filter lists.
For each domain, we extract its  category, owners and country of registration

Table~\ref{tab:ELEPusage} summarizes the usage of \ELEP\ lists in  the previous works 
that we describe in Section~\ref{sec:relatedwork}.

\begin{table*}[!th]
\SHORTEN
\caption{Usage of \ELEP\ lists in security, privacy, and web measurement communities (venues from 2016-2018).
``Detection'' describes how \ELEP\ was used to detect trackers: whether the filterlists were applied 
on all requests, (``Req''), 
on requests and follow-up requests that would be blocked,  (``Req.+Follow'') 
or whether filterlists were further customised before being applied to the dataset (``Custom''). 
In the dependency column, ``Rely'' means that the authors use the {\ELEP} to build their results, ``verify'' means that the authors only use \ELEP lists to verify their results.
}
\label{tab:ELEPusage}
\centering
\begin{tabular}{p{3.9cm} p{2.9cm} p{1.5cm} p{1.5cm} p{1.5cm} p{1.5cm}  p{1.0cm} }
\hline 
Paper & Venue	 						& 	EasyList & EasyPrivacy		&{Detection} 	& \multicolumn{1}{c}{Dependency} 	 \\
 		\hline
Englehardt and Narayanan~\cite{Engl-Nara-16-CCS} & 
	ACM CCS 2016 						&  \checkmark & \checkmark   	& Req. 		 	&  Rely \\
Bashir et al.~\cite{Bash-etal-16-USENIX} & 
	USENIX Security 2016				& \checkmark &   				& Custom. 		&   Rely \\
Lauinger et al. \cite{Laui-etal-17-NDSS} & 
	NDSS 2017 							& \checkmark &\checkmark  		& Req.+Follow 		&  Rely \\
Razaghpanah et al.~\cite{Raza-etal-18-NDSS} & 
	NDSS 2018  						& \checkmark & 					& Custom. 		& Rely \\
Ikram et al.~\cite{Ikra-etal-17-PETS} & 
	PETs 2017 							& \checkmark & 					& Req.+Follow 		& Verif. \\
Englehardt et al.\cite{Engl-etal-18-PETS} & 
	PETs 2018  							& \checkmark & \checkmark  		& Req.+Follow 			&  Verif. \\	
Bashir and Wilson~\cite{Bash-Wils-18-PETS} & 
	PETs 2018							& \checkmark & \checkmark   	& Custom. 		& Rely+Verif.\\
Bashir et al.\cite{Bash-etal-18-IMC} 	& 	
	IMC 2018 							& \checkmark & \checkmark   	& Custom. 		& Rely		\\%
Iordanou \etal \cite{Iord-etal-18-IMC} & 
	IMC 2018 							&  \checkmark & \checkmark 		& Req.+Follow		&  Rely \\
\hline
\end{tabular}
\label{tab:ELEPDinliterature}
\end{table*}